

\magnification=\magstep1
\hsize=15.7truecm \vsize=23.4truecm
\baselineskip=6mm
\font\bg=cmbx10 scaled 1200
\footline={\hfill\ -- \folio\ -- \hfill}
\def\prenum#1{\rightline{#1}}
\def\date#1{\rightline{#1}}
\def\title#1{\centerline{\bg#1} \vskip 5mm}
\def\author#1{\centerline{#1} \vskip 3mm}
\def\address#1{\centerline{\sl#1}}
\def\abstract#1{{\centerline{\bg Abstract}} \vskip 3mm \par #1}
\def\acknowledgments{{\centerline{\bg Acknowledgement}} \vskip 2mm}
\def\references{{\centerline{\bg References}} \vskip 3mm}
\def\table{{\centerline{\bg Table Captions}} \vskip 5mm}
\def\chapter#1{\centerline{\bg#1} \vskip 5mm}
\def\section#1{\centerline{\bg#1} \vskip 5mm}
\def\endpage{\vfill \eject}
\def\lq{\char"5C}

\def\footnotes#1#2{\baselineskip=5truemm \footnote{#1}{#2} \baselineskip
=6mm}
\fontdimen5\textfont2=1.2pt

\def\semiprod{ \ \mathrel \times \mathrel{\mkern-4mu} \mathrel
{\vrule height4.7pt width .2pt depth0pt \ } }

\prenum{NBI--HE--92--34}
\prenum{KOBE--92--04}
\date{June 1992}

\vskip 10mm

\title{Symmetries between Untwisted and Twisted Strings}
\title{on Asymmetric Orbifolds}
\vskip 20mm

\author{Y. IMAMURA${^a}$, M. SAKAMOTO${^b}$\footnotes{*}{On leave from
Department of Physics, Kobe University, Nada, Kobe 657, Japan}
, T. SASADA${^a}$ and M. TABUSE${^c}$}
\address{${^a}$ Department of Physics, Kobe University}
\address{Nada, Kobe 657, Japan}

\vskip 3mm

\address{${^b}$ The Niels Bohr Institute, University of Copenhagen}
\address{Blegdamsvej 17, DK-2100 Copenhagen X, Denmark}

\vskip 3mm

\address{${^c}$ Department of Computer Science and Systems Engineering,}
\address{Miyazaki University, Miyazaki 889-21, Japan}

\vfill

\abstract
We study symmetries between untwisted and twisted strings on asymmetric
orbifolds. We present a list of asymmetric orbifold models to possess
intertwining currents which convert untwisted string states to twisted
ones, and  vice versa.
We also present a list of heterotic strings on asymmetric orbifolds with
supersymmetry between untwisted and twisted string states. Some of
properties inherent in asymmetric orbifolds, which are not shared by
symmetric orbifolds, are pointed out.

\endpage

\section{1. Introduction}

In the construction of realistic four-dimensional string models, various
approaches have been proposed [1-8]. Among them, the orbifold
compactification [1] is probably the most efficient method and the
orbifold compactification of the heterotic string [9] is believed to
provide a phenomenologically realistic string model.
The heterotic string has asymmetric nature: The left-movers consist of a
26-dimensional bosonic string and the right-movers consist of a
10-dimensional superstring. This asymmetric nature of the heterotic string
naturally leads to the idea of asymmetric orbifolds [10].
Although the search for realistic orbifold models has been continued by
many authors [11-14], a more general and systematic investigation of
asymmetric orbifolds should be done. In this paper we shall reveal some of
properties inherent in asymmetric orbifolds, which are not shared by
symmetric orbifolds.

Suppose that there exists an intertwining current operator which converts
string states in an untwisted sector to string states in a twisted sector
in an asymmetric orbifold model. This current operator will correspond to
a state of the conformal weight (1,0) (or (0,1)) in the twisted sector and
connect the ground state of the untwisted sector to the (1,0) (or (0,1))
twisted state.
Therefore, the existence of a (1,0) (or (0,1)) twisted state implies a
symmetry between the untwisted and twisted sectors. Since a total Hilbert
space of strings on the orbifold is a direct sum of the untwisted and
twisted Hilbert spaces, the existence of (1,0) (or (0,1)) physical twisted
states implies that the symmetry of the total Hilbert space is larger than
the symmetry of each (untwisted or twisted) Hilbert space
{\footnotes{$^\star$}{\ Some\  examples\  have\  been\  discussed\  in\
refs.\ [1,15,16]\  and\  in\  our\  previous\  papers\ [17,18].}}.
It should be emphasized that this symmetry \lq enhancement" does not occur
in the case of symmetric orbifolds because the left- and right- conformal
weights, $h$ and $\bar h$, of a ground state of any twisted sectors are
both positive (and equal) for symmetric orbifolds and hence no (1,0) (or
(0,1)) state appears in any twisted sector.

     Symmetry \lq enhancement" stated above will mean \lq enhancement" of
gauge symmetries. In the case of superstring theories, another interesting
type of symmetry \lq enhancement" might occur, i.e., supersymmetry \lq
enhancement". N=1 space-time supersymmetry might appear in a spectrum of a
total Hilbert space through supersymmetry \lq enhancement" even though
there is no unbroken space-time supersymmetry in each untwisted or twisted
Hilbert space. It would be of interest to investigate such a new class of
four-dimensional orbifold models with N=1 space-time supersymmetry.

      \ Another\  peculiarity\  of\  asymmetric\  orbifolds\  is\  related
to the \lq torus-orbifold equivalence" [1,3,19-21]. An orbifold will be
obtained by dividing a ($D$-dimensional) torus $T^D$ by the action of a
discrete symmetry group $P$ of the torus. We may denote the orbifold by
$T^D/P$. An orbifold model with $T^D/P$ can be equivalent to a torus model
with $T'^D$ for asymmetric as well as symmetric orbifolds. However, $T^D$
can be equal to $T'^D$ only for asymmetric orbifolds but not symmetric
ones. If the symmetry group $P$ includes an outer automorphism of the
lattice defining the torus, then the orbifold model cannot be rewritten as
a torus model for symmetric orbifolds.
Our results, however, suggest a new class of the \lq torus-orbifold
equivalence", that is, some asymmetric orbifold models can be rewritten as
torus models even in the case of outer automorphisms.

   In the next section, we discuss general properties of asymmetric ${\bf
Z}_N$-orbifold models, which do not depend on specific momentum lattices
on which left- and right-moving momenta lie. In sect. 3, we investigate ${
\bf Z}_N$-automorphisms of Lie algebra lattices. We will be concerned with
${\bf Z}_N$-automorphisms which have no fixed direction and give a
classification of momentum lattices associated with Lie algebras and their
${\bf Z}_N$-automorphisms.
In sect. 4, we briefly review the \lq torus-orbifold equivalence", which
may be used to determine full symmetries of asymmetric orbifold models. In
sect. 5, we present a list of asymmetric ${\bf Z}_N$-orbifold models which
possess (1,0) twisted states and show that these states correspond to
twist-untwist intertwining currents which convert untwisted string states
to twisted string states, and vise versa. In sect. 6, we study $E_8 \times
E_8$ heterotic strings on asymmetric ${\bf Z}_N$-orbifolds and present
four orbifold models with supersymmetry between untwisted and twisted
string states.
In sect. 7, various properties inherent in asymmetric orbifold models are
summarized. In an appendix, shift vectors which are introduced in
rewriting orbifold models into torus models are given.

\vskip 10mm

\section{2. Asymmetric ${\bf Z}_N$-orbifolds}

An orbifold [1] will be obtained by dividing a torus by the action of a
discrete symmetry group $P$ of the torus. In the construction of an
orbifold model, we start with a $D$-dimensional toroidally compactified
closed bosonic string theory which is specified by a $(D+D)$-dimensional
lorentzian even self-dual lattice $\Gamma^{D,D}$ [22]. The left- and
right-moving momentum $(p_L^I,p_R^I)\ (I=1, \dots ,D)$ lies on the lattice
$\Gamma^{D,D}$. Let $g$ be a
 group element of $P$. The $g$ will, in general, act on the left-movers
and the right-movers differently. If both the left- and right-moving
string coordinates $X_L^I$ and $X_R^I$ obey twisted boundary conditions,
there are no (1,0) (or (0,1)) states in the twisted sector because the
conformal weight $(h,\bar h)$ of the ground state in the twisted sector
will be positive, i.e., $h>0$ and $\bar h>0$.
For (1,0) ((0,1)) states to appear in a twisted sector, the right- (left-)
moving string coordinate must obey the untwisted boundary condition. Hence
we will restrict our considerations to the following class of the ${\bf
Z}_N$-transformation:
$$ g: (X_L^I,X_R^I) \rightarrow (U^{IJ}X_L^J,X_R^I), \quad (I,J=1, \dots ,
D), \eqno(2.1)$$
where $U$ is a rotation matrix which satisfies $U^N=\bf 1$. The ${\bf
Z}_N$-transformation must be an automorphism of the lattice $
\Gamma^{D,D}$, i.e.,
$$ (U^{IJ}p_L^J,p_R^I) \in \Gamma^{D,D} \quad {\rm for \ all} \quad
(p_L^I,p_R^I) \in \Gamma^{D,D}. \eqno(2.2)$$
For simplicity, we will assume that the rotation matrix $U$ and all its
powers $U^{\ell}\ (\ell=1, \dots ,N-1)$ do not have any fixed direction.

Let us consider the $g^{\ell}$-twisted sector in which strings close up to
the $g^{\ell}$-action. The one loop partition function of the $g^{
\ell}$-sector twisted by $g^{m}$ is given by
$$ Z(g^\ell, g^m; \tau)={\rm Tr}[g^m e^{i2\pi\tau(L_0-{D\over 24})} e^{-i2
\pi\bar \tau ({\bar L}_0-{D\over 24})}], \eqno(2.3)$$
where the trace is taken over the Hilbert space of the $g^\ell$-sector and
$L_0$ $(\bar L_0)$ is the zero mode of the left- (right-) moving Virasoro
operators. Modular invariance of the one loop partition function will
require
$$ Z(g^\ell,g^m;\tau+1)=Z(g^\ell,g^{m+\ell};\tau), \eqno(2.4)$$
$$ Z(g^\ell,g^m;-1/\tau)=Z(g^{-m},g^{\ell};\tau). \eqno(2.5)$$
Let $N_\ell$ be the minimum positive integer such that $(g^\ell)^{N_\ell}$
=1. Since $g^N$=1 and hence $Z(g^\ell,g^m;\tau)$ has to be invariant under
the modular transformation $\tau \rightarrow \tau + N_\ell$, the necessary
condition for modular invariance is
$$ N_\ell(L_0-\bar L_0)=0 \quad {\rm mod} \  1. \eqno(2.6)$$
This is called the left-right level matching condition  and it has been
proved that this condition is also sufficient for modular invariance
[10,23]. Let $\Gamma_0$ be the $g^\ell$-invariant sublattice of $
\Gamma^{D,D}$. Then the left-and right-moving momentum in the $g^
\ell$-sector lies on the lattice ${\Gamma_0}^*$ [10], which is the dual
lattice of $\Gamma_0$ {\footnotes{$^\star$}{This is not true in general,
as pointed out in refs. [24,25]. We will, however, restrict our
considerations to the models in which the left- and right-moving momentum
in the $g^\ell$-sector lies on the lattice ${\Gamma_0}^*$.}}. The degeneracy
$d_\ell$ of the ground states in the $g^\ell$-sector is given by [10]
$$ d_\ell={\sqrt{{\rm det}'({\bf 1}-U^\ell)} \over V_{\Gamma_0}}, \eqno
(2.7)$$
where $V_{\Gamma_0}$ is the volume of the unit cell of the lattice $
\Gamma_0$ and the determinant is taken over the eigenvalues of $U^\ell$
not equal to one. The left-right level matching condition (2.6) can
equivalently be rewritten as the following two conditions:
$$ N_\ell(h_\ell-\bar h_\ell) = 0 \quad {\rm mod}\  1, \eqno(2.8)$$
$$ N_\ell( ({p_L^I})^2 - ({p_R^I})^2 ) = 0 \quad {\rm mod}\ 2  \qquad  {
\rm for \  all}\  (p_L^I,p_R^I) \in {\Gamma_0}^* , \eqno(2.9)$$
where $h_\ell\ (\bar h_\ell)$ denotes the conformal weight of the ground
state in the $g^\ell$-sector with respect to the left- (right-) movers.
Since the ${\bf Z}_N$-transformation is given by eq. (2.1) and $U^\ell\ (
\ell=1, \dots , N-1)$ is assumed to have no fixed direction, the two
conditions (2.8) and (2.9) reduce to
$$ N_\ell h_\ell = 0 \quad {\rm mod}\  1, \eqno(2.10)$$
$$ N_\ell ({p_R^I})^2 = 0 \quad {\rm mod}\ 2  \qquad  {\rm for \  all}\
p_R^I \in {\Gamma_0}^* .\eqno(2.11)$$
Let $\omega^{s_a}\ (a=1, \dots, D)$ be an eigenvalue of $U^\ell$ with $0
\leq s_a \leq N-1$, where $\omega=e^{i2\pi/N}$. Then $h_\ell$ is given by
$$ h_\ell={1\over4}\sum_{a=1}^D{s_a\over N} \Bigl( 1-{s_a\over N} \Bigr).
\eqno(2.12)$$
{}From eqs. (2.3) and (2.4), it turns out that the operator $g^\ell$ in the
$g^\ell$-sector is given by
$$ g^\ell ={\rm exp}[i2\pi(L_0-\bar L_0)]. \eqno(2.13)$$
If $\ell$ is relatively prime to $N$, $g^\ell=1$ means $g=1$. Thus any
state of $L_0-\bar L_0=0$\ mod 1, in particular, $(L_0,\bar L_0)=(1,0)$ is
${\bf Z}_N$-invariant and hence physical. If $\ell$ is not relatively
prime to $N$, $g^\ell =1$ does not mean $g=1$. Thus, to determine a
physical spectrum of the $g^\ell$-sector, a detailed analysis of the one
loop partition function is required.

   Let $\omega^{r_a}\ (a=1, \dots , D)$ be an eigenvalue of $U$, where $
\omega=e^{i2\pi/N}$. Since we have assumed that the rotation matrix $U$
and all its powers $U^\ell\ (\ell=1, \dots, N-1)$ have no fixed direction,
$\omega^{\ell r_a}$ is not equal to one for $\ell=1, \dots , N-1$. This
implies that $\omega^{r_a}\ (a=1, \dots, D)$ must be a primitive $N$th
root of unity. Every primitive $N$th root of unity can be written as $
\omega^m\ (m=1, \dots, N-1)$, where $m$ is relatively prime to $N$. The
number of the primitive $N$th roots of unity is denoted by $\varphi (N)$,
which is called the Euler function.
If we change the $(D+D)$-dimensional basis vectors to the lattice basis of
$\Gamma^{D,D}$, then the ${\bf Z}_N$-automorphism $g$ in eq. (2.1) is
represented by an integer matrix. This means that the characteristic
polynomial det$(\lambda{\bf 1}-U)$ must have integer coefficients.
It turns out
that det$(\lambda{\bf 1}-U)$ is given by a multiple of the cyclotomic
polynomial $\Phi_N(\lambda)$ [26], i.e.,
$$ {\rm det}(\lambda{\bf 1}-U)=[\Phi_N(\lambda)]^{D/{\varphi(N)}}. \eqno
(2.14) $$
In eq. (2.14), $\Phi_N(\lambda)$ is the polynomial of the degree $\varphi
(N)$ and is defined by
$$ \Phi_N(\lambda)=
\prod_{\scriptstyle (m,N)=1
\atop
\scriptstyle m=1,\dots,N-1}
(\lambda-\omega^m), \eqno(2.15) $$
where $(m,N)$ denotes the greatest common divisor of $m$ and $N$.
Therefore, we have found that the eigenvalues of $U$ consist of the
primitive $N$th roots of unity and that each primitive $N$th root of unity
appears $D/{\varphi(N)}$ times. This implies that the dimension $D$ must
be a multiple of $\varphi(N)$, i.e.,
$$ D=0 \quad {\rm mod}\  \varphi(N). \eqno(2.16) $$
A list of $\varphi(N)$ is given in table 1.

We can further put some constraints on the dimension $D$, irrespective of
the specific lattice $\Gamma^{D,D}$. Since $U^\ell\ (\ell=1, \dots , N-1)$
is assumed to have no fixed direction, we have
$$ {\rm det}(\lambda{\bf 1}-U^\ell)=[\Phi_{N_\ell} (\lambda)]^{D/{\varphi
(N_\ell)}}, \eqno(2.17) $$
where $N_\ell$ is the minimum positive integer such that $(g^\ell)^{N_
\ell}=1$. It follows from the formula (2.12) that
$$ h_\ell={D\over {\varphi(N_\ell)}}{1\over4}
\sum_{\scriptstyle (m,N_\ell)=1
\atop
\scriptstyle m=1,\dots,N_\ell-1}
{m\over N_\ell} \Bigl( 1-{m\over N_\ell}\Bigr).  \eqno(2.18)$$
Since modular invariance requires eq. (2.10), the dimension $D$ must
satisfy
$$ D=0 \quad {\rm mod} \ {\varphi(N_\ell)\over {N_\ell h'_\ell}} \quad {
\rm for}\ \ell=1, \dots , N-1, \eqno(2.19) $$
where
$$ h'_\ell={1\over4}
\sum_{\scriptstyle (m,N_\ell)=1
\atop
\scriptstyle m=1,\dots,N_\ell-1}
{m\over N_\ell} \Bigl( 1-{m\over N_\ell}\Bigr).  \eqno(2.20)$$
For example, $D$ must be a multiple of 8 for $N=2$ because $\varphi(2)=1$
and $h'_1={1\over 16}$. In the third column of table 1, allowed values of
$D$, which are consistent with the constraints (2.16) and (2.19), are
listed up to $N=30$.

   Since we are interested in asymmetric orbifold models which possess
(1,0) states in the twisted sectors, $h_\ell\ (\ell=1, \dots , N-1)$ must
be less than or equal to one for some $\ell$. This requirement severely
restricts allowed values of $D$. Since $h_\ell$ is proportional to $D$,
$h_\ell$ will exceed one for appropriately large $D$. Therefore, only
lower dimensional orbifold models might possess (1,0) states in twisted
sectors. In the fourth column of table 1, a list of the dimensions $D$ in
which (1,0) states might appear in twisted sectors is given up to $N=30$.
In sect. 5, we will see all asymmetric orbifold models listed in the forth
column of table 1, which possess (1,0) states in twisted sectors, except
for $N=14$, 21, 25 and 26.

\vskip 10mm

\section{3. Automorphisms of $\Gamma^{D,D}$}

In this section we will investigate automorphisms of the momentum lattice
$\Gamma^{D,D}$. Since a complete classification of automorphisms of
general
lorentzian even self-dual lattices is not known, we will restrict our
considerations to lattices associated with Lie algebras.

We take the lattice $\Gamma^{D,D}$ of an asymmetric ${\bf Z}_N$-orbifold
to be of
the form:
$$
\Gamma^{D,D}=\{\ (p^I_L,p^I_R)\ \vert \ p^I_L,p^I_R\in \Lambda^\ast ,\
p^I_L-p^I_R
\in \Lambda\ \},
\eqno(3.1)
$$
where $\Lambda$ is a $D$-dimensional lattice and $\Lambda^\ast$ is the
dual
lattice of $\Lambda$. It turns out that $\Gamma^{D,D}$ is lorentzian even
self-dual if $\Lambda$ is even integral. In the following, we will take
$\Lambda$ in eq. (3.1) to be a root lattice of a simply-laced semi-simple
Lie
algebra $G$ [27]. Then the lattice $\Lambda$ is even integral if the
squared
length of the root vectors is normalized to $2m\ (m=1,2,$\dots$)$.
Since the ${\bf Z}_N$-transformation defined in eq. (2.1) must be an
automorphism of
the lattice (3.1), the rotation matrix $U$ must be an automorphism of
$\Lambda$ as well as $\Lambda^\ast$. Furthermore, the condition (2.2)
requires $U$ to satisfy
$$
U^{IJ}p^J_L-p^I_L\in \Lambda \ \ \ {\rm for\ all} \ p^I_L\in \Lambda^\ast
{}.
\eqno(3.2)
$$

Let us first consider the case that the squared length of the root vectors
in
the root lattice $\Lambda$ is normalized to two. Then the dual lattice
$\Lambda^\ast$ is equivalent to the weight lattice of the Lie algebra $G$.
For
simplicity, we assume that the rotation matrix $U$ and all its powers
$U^\ell \ (\ell =1,\dots ,N-1)$ have no fixed direction.
Let $\Phi$ be a root system of the semi-simple Lie algebra $G$ and let
Aut$\Phi$ be the group of automorphisms of this root system. The group
Aut$\Phi$ is a semi-direct product of two groups [28]:
$$
{\rm Aut}\Phi=W\semiprod {\rm Aut}(\Phi,\Delta),
$$
$$W\cap {\rm Aut}(\Phi,\Delta)=\{1\},
\eqno(3.3)
$$
where $W$ is the Weyl group of $\Phi$, which is a normal subgroup of Aut$
\Phi$, and Aut$(\Phi,\Delta)$ is defined as
$$
{\rm Aut}(\Phi,\Delta)=\{\ \varphi\in {\rm Aut}\Phi\ \vert\
\varphi(\Delta)=\Delta\ \}.
\eqno(3.4)
$$
Here, $\Delta$ is a fixed basis of $\Phi$. Aut$(\Phi,\Delta)$ corresponds
to
the group of symmetries of the Dynkin diagram of $G$. The condition (3.2)
tells us that the automorphism $U$ of $\Lambda$ must not change a
conjugacy
class of any vector in $\Lambda^\ast$. Since the squared length of the
root
vectors is normalized to two, any element of the Weyl group $W$ in eq.
(3.3)
does not change a conjugacy class of any vector in $\Lambda^\ast$. On the
other hand, for any nontrivial element of
Aut$(\Phi,\Delta)$ in eq. (3.3),
there always exists a vector in $\Lambda^\ast$ which is mapped to a
different
conjugacy class of $\Lambda^\ast$ by the automorphism. Consequently, the
automorphism $U$ of $\Lambda$ must be an element of the Weyl group $W$ of
the root system~$\Phi$.

All automorphisms of the root systems of simple Lie algebras such that
they
and all their powers have no fixed directions have been discussed in
ref. [29]. In the following, we will give the result concerning the
Weyl group elements of the root system of simply-laced simple Lie
algebras.
In the case of $G=SU(n+1)$,
there exists a Weyl group element of $SU(n+1)$
which has no fixed directions if and only if $n+1$ is prime. The order of
this
Weyl group element is $n+1$. In the case of $G=SO(2n)$, there exist Weyl
group
elements of $SO(2n)$ which have no fixed directions if and only if $n$ is
even, i.e., $n=2^{\ell}p\ (\ell>0, p={\rm odd})$. Then the allowed orders
are
given by $2,2^2,\dots,2^{\ell}$.
In the cases of $G=E_6,E_7 \ {\rm and} \ E_8$,
the orders of the Weyl group elements which have no fixed directions are

\noindent \hskip 4.5cm $E_6:3,9,$

\noindent \hskip 4.5cm $E_7:2,$

\noindent \hskip 4.5cm $E_8:2,3,4,5,6,8,10,12,15,20,24,30.$

\noindent
The explicit expressions of the Weyl group elements stated above are given
in
ref. [29].

Let us next consider the case that the squared length of the root vectors
in
$\Lambda$ is normalized to $2m\ (m=2,3,\dots)$. Then we have
$\Lambda=\sqrt{m}\Lambda_R$ and
$\Lambda^{\ast}={1\over \sqrt m}\Lambda_W\ (m=2,3,\dots)$, where $
\Lambda_R$
denotes the root lattice of $G$ in which the squared length of the root
vectors is normalized to two and $\Lambda_W={\Lambda_R}^\ast$. Let
$\mu^i\ (i=1,\dots,{\rm rank}G)$ be a fundamental weight of $G$. Since
$\Lambda_R$ is an even lattice, the condition (3.2) implies that
$$
(\mu^i)^2-\mu^i\cdot U\mu^i=0\ \ {\rm mod}\ m^2\ \ \ {\rm for}\ i=1,
\dots,{\rm rank}G.
\eqno(3.5)
$$
Since $U$ is a rotation matrix which has no fixed directions, the left
hand
side of eq. (3.5) is restricted to
$$
0<(\mu^i)^2-\mu^i\cdot U\mu^i\leq2(\mu^i)^2,
\eqno(3.6)
$$
where the equality holds if and only if $U\mu^i=-\mu^i$. It is known that
there always exists a fundamental weight $\mu^i$ such that $(\mu^i)<2$
except
for $G=(E_8)^\ell\ (\ell=1,2,\dots)$ [30]. For $G=(E_8)^\ell\ (\ell=1,2,
\dots)$,
all fundamental weights have the squared length two because the root
lattice
of $G=(E_8)^\ell$ is even self-dual. Hence the conditions (3.5) are
satisfied
if and only if $m=2,\ U=-{\bf 1}$ and $G=(E_8)^\ell\ (\ell=1,2,\dots)$.

We have investigated the allowed automorphisms of the lattice (3.1)
associated
with Lie algebra lattices $\Lambda$ and have found that the orders $N$ of
the
allowed automorphisms are given by $N={\rm prime \ numbers}$,
$2^M\ (M=1,2,\dots)$ and 6,9,10,12,15,20,24,30. Since we are
interested in asymmetric orbifold models
which possess (1,0) states in twisted sectors, the allowed orders further
reduce as follows: If the order $N$ is prime, the left conformal weight of
the
ground states of every $g^\ell$-twisted sector $(\ell=1,\dots,N-1)$ has
the
common value, i.e.,
$$
h_\ell
={1\over 4}\sum^{N-1}_{i=1}{i\over N}\Bigl(1-{i\over N}\Bigr){D\over N-1}
={N+1\over 24N}D.
\eqno(3.7)
$$
It follows that the existence of (1,0) states in some twisted sectors
implies
$D<24$. Thereby, the allowed prime orders are $N$=2,3,5,7,11,13,17,19,23.
If
the order $N$ is $2^M\ (M=1,2,\dots)$, the left conformal weight of the
ground
states of the $g^\ell$-twisted sector $(\ell=1,\dots,N-1)$ is given by
$$
h_\ell
={1\over 4}\sum^{N_\ell/2}_{j=1}{2j-1\over N_\ell}
\Bigl(1-{2j-1\over N_\ell}\Bigr){D\over N_\ell/2}
={(N_\ell)^2+2\over 24(N_\ell)^2}D.
\eqno(3.8)
$$
It follows that the existence of (1,0) states in some twisted sectors
implies
$D<24$. Thereby, the allowed orders of the form $2^M\ (M=1,2,\dots)$ are
$N$=
2,4,8,16.

\vskip 10mm

\section{4. Torus-Orbifold Equivalence}

In the next section, we will present asymmetric ${\bf Z}_N$-orbifold
models with intertwining currents which convert untwisted string states to
twisted ones, and vice versa. To investigate symmetries of those orbifold
models, we might construct the intertwining currents explicitly. Such
currents correspond to twisted state emission vertex operators with the
conformal weight (1,0). It is not, however, easy to explicitly construct
twisted state emission vertex operators [15,16,31-33].
To avoid this difficulty, we will use a trick, the \lq torus-orbifold
equivalence" [1,3,19-21]: Any closed bosonic string theory compactified on
a ${\bf Z}_N$-orbifold is equivalent to a closed bosonic string theory on
a torus if the dimension of the orbifold is equal to rank of a gauge
symmetry of strings in each of the untwisted and twisted sectors of the
orbifold model, or equivalently, if the ${\bf Z}_N$-transformation is an
inner automorphism of the lattice $\Gamma^{D,D}$.
In the next section, we may rewrite orbifold models into equivalent torus
models using the \lq torus-orbifold equivalence" and investigate full
symmetries of the torus models instead of the orbifold models themselves.
For later convenience, we will give a brief review of the \lq
torus-orbifold equivalence" in this section. See ref. [21] for details.

Let us start with a $D$-dimensional torus model associated with the root
lattice $\Lambda_R(G)$ of a simply-laced Lie algebra $G$\ ($D$= rank $G$).
Suppose that an affine Kac-Moody algebra $\hat G \oplus \hat G$ is
realized in the vertex operator representation {\sl ${\grave a}$ la}
Frenkel-Kac and Segal [34]:
$$\eqalign{ P^I_L(z)&\equiv i\partial_z X^I_L(z),\cr
     V_L(\alpha ; z)&\equiv : {\rm exp}\{i\alpha \cdot X_L(z)\} : \ , \cr}
\eqno(4.1)$$
and
$$\eqalign{ P^I_R(\bar z)&\equiv i\partial_{\bar z} X^I_R(\bar z),\cr
     V_R(\alpha ; \bar z)&\equiv : {\rm exp}\{i\alpha \cdot X_R(\bar z)\}
: \ ,\cr}\eqno(4.2)$$
where $\alpha$ is a root vector of $G$ and its squared length is
normalized to two. A ${\bf Z}_N$-orbifold model is obtained by modding out
of this torus model by a ${\bf Z}_N$-rotation which is an automorphism of
the lattice defining the torus. Since every physical string state on the
${\bf Z}_N$-orbifold must be invariant under the ${\bf
Z}_N$-transformation, the ${\bf Z}_N$-invariant subgroup $G_0$ of $G$ is
the unbroken gauge symmetry in the untwisted sector.

Let us consider the $g^\ell$-sector (the untwisted sector for $\ell=0$ and
the twisted sectors for $\ell=1, \dots , N-1$). As eqs. (4.1) and (4.2),
we will have the operators $P^I_L(z)$, $V_L(\alpha ; z)$ and $P^I_R(\bar
z)$, $V_R(\alpha ; \bar z)$ which generate the untwisted (for $\ell=0$) or
twisted (for $\ell=1, \dots , N-1$) affine Kac-Moody algebra $\hat G
\oplus \hat G$. If rank of $G_0$ is equal to $D$, we can always construct
the ${\bf Z}_N$-invariant operators $P'^I_L(z)$ and $P'^I_R(\bar z)$ ($I
=1, \dots , D$) from suitable linear combinations of $P^I_L(z)$, $V_L(
\alpha ; z)$ and $P^I_R(\bar z)$, $V_R(\alpha ; \bar z)$ such that
$$ g(P'^I_L(z), P'^I_R(\bar z))g^{-1} = (P'^I_L(z), P'^I_R(\bar z)) ,
\eqno(4.3)$$
and
$$ P'^I_L(w)P'^J_L(z)={\delta^{IJ}\over (w-z)^2}+({\rm regular \ \
terms}), $$
$$ P'^I_R(\bar w)P'^J_R(\bar z)={\delta^{IJ}\over (\bar w-\bar z)^2}+({\rm
regular\ \  terms}),\eqno(4.4) $$
where $g$ is the operator which generates the ${\bf Z}_N$-transformation
in the $g^\ell$-sector. It follows from (4.4) that $P'^I_L(z)$ and $P'^I_R
(\bar z)$ can be expanded as
$$ P'^I_L(z) \equiv i\partial_z X'^I_L(z) \equiv \sum_{n\in Z}\alpha'^I_{L
n}z^{-n-1},$$
$$ P'^I_R(\bar z) \equiv i\partial_{\bar z} X'^I_R(\bar z) \equiv \sum_{n
\in Z}\alpha'^I_{R n}{\bar z}^{-n-1} , \eqno(4.5)$$
with
$$[\alpha'^I_{Lm},\alpha'^J_{Ln}]=m\delta^{IJ}\delta_{m+n,0},$$
$$[\alpha'^I_{Rm},\alpha'^J_{Rn}]=m\delta^{IJ}\delta_{m+n,0}. \eqno(4.6)$$
In this basis, a vertex operator in the $g^\ell$-sector will be given by
$$ V'(k_L, k_R ; z)=\ ^\circ_\circ {\rm exp}\{ik_L\cdot X'_L(z)+ik_R
\cdot X'_R(\bar z)\}^\circ_\circ, \quad{\rm for}\ (k_L^I, k_R^I) \in
\Gamma^{D,D},\eqno(4.7)$$
where $\ ^\circ_\circ \ ^\circ_\circ$ denotes the normal ordering with
respect to the new basis of the operators. Since $P'^I_L(z)$ and $P'^I_R(
\bar z)$ are invariant under the ${\bf Z}_N$-transformation, the vertex
operator (4.7) will transform as
$$ gV'(k_L,k_R;z)g^{-1} = e^{i2\pi(k_L \cdot v_L-k_R \cdot v_R)}V'
(k_L,k_R;z), \eqno(4.8)$$
for some constant vector ($v^I_L,v^I_R$). It follows that $g$ will be
given by
$$ g=\eta_{(\ell)}{\rm exp}\{i2\pi(p'_L\cdot v_L - p'_R\cdot v_R)\}, \eqno
(4.9)$$
where $p'^I_L=\alpha'^I_{L0}$, $p'^I_R=\alpha'^I_{R0}$ and $\eta_{(\ell)}$
is a constant phase with $(\eta_{(\ell)})^N=1$. Thus, the string
coordinate in the new basis transforms as
$$ g(X'^I_L(z), X'^I_R(\bar z))g^{-1} = (X'^I_L(z)+2\pi v^I_L, X'^I_R(\bar
z)-2\pi v^I_R). \eqno(4.10)$$
This implies that the string coordinate $(X'^I_L(z), X'^I_R(\bar z))$ in
the $g^\ell$-sector obeys the following boundary condition:
$$ (X'^I_L(e^{2\pi i}z), X'^I_R(e^{-2\pi i}\bar z)) = (X'^I_L(z)+2\pi \ell
v^I_L, X'^I_R(\bar z)-2\pi \ell v^I_R) + ({\rm torus \ \ shift}), \eqno
(4.11)$$
and hence that the eigenvalues of the momentum $(p'^I_L,p'^I_R)$ in the
new basis are of the form
$$ (p'^I_L, p'^I_R) \in \Gamma^{D,D} + \ell (v^I_L, v^I_R). \eqno(4.12)$$
In the new basis, $g^\ell$ in the $g^\ell$-sector will be given by
$$ g^\ell=e^{i2\pi(L'_0-{\bar L}'_0)}, \eqno(4.13)$$
where
$$ L'_0=\sum_{I=1}^D\{{1\over 2}(p'^I_L)^2+\sum_{n=1}^\infty\alpha'^I_{L
-n}\alpha'^I_{L n}\}, $$
$$ {\bar L}'_0=\sum_{I=1}^D\{{1\over 2}(p'^I_R)^2+\sum_{n=1}^\infty
\alpha'^I_{R -n}\alpha'^I_{R n}\}. \eqno(4.14)$$
Comparing eq. (4.13) with eq. (4.9), we find
$$ \eta_{(\ell)}={\rm exp}\{-i\pi\ell((v^I_L)^2-(v^I_R)^2)\}. \eqno
(4.15)$$
Since $g^N = 1$, $(v^I_L, v^I_R)$ must satisfy
$$ N((v^I_L)^2-(v^I_R)^2)=0 \quad {\rm mod}\ 2, $$
$$ N(v^I_L, v^I_R) \in \Gamma^{D,D}. \eqno(4.16)$$
Every physical state must obey the condition $g=1$ because it must be
invariant under the ${\bf Z}_N$-transformation. Thus the allowed momentum
eigenvalues $(p'^I_L,p'^I_R)$ of the physical states in the $g^
\ell$-sector are restricted to
$$ (p'^I_L, p'^I_R) \in \Gamma^{D,D} + \ell (v^I_L, v^I_R) \ \  {\rm with
}\ \  p'_L\cdot v_L-p'_R\cdot v_R -{1\over 2} \ell ((v^I_L)^2-(v^I_R)^2)=0
\ \ {\rm mod}\ 1. \eqno(4.17)$$

The total physical Hilbert space $\cal H$ of the ${\bf Z}_N$-orbifold
model is a direct sum of a physical space $\cal H_{(\ell)}$ in each
sector:
$$ {\cal H} = {\cal H}_{(0)}\oplus{\cal H}_{(1)}\oplus \cdots \oplus{\cal
H}_{(N-1)}. \eqno(4.18)$$
In the above consideration we have shown that $\cal H$ is equivalent to
$$ {\cal H} = \{\alpha'^I_{L-m}\cdots\alpha'^J_{R-n}\cdots |
{p'}_L^I,{p'}_R^I> |m, \dots , n,\dots \in {\bf Z} > 0,\
({p'}_L^I,{p'}_R^I)\in {\Gamma'}^{D,D}\},\eqno(4.19)$$
where
$$ {\Gamma'}^{D,D}=\{({p'}^I_L,{p'}^I_R)\in \bigcup_{\ell=0}^{N-1}[
\Gamma^{D,D}+\ell(v^I_L,v^I_R)]\ |\ p'_L\cdot v_L - p'_R\cdot v_R -{1\over
2}\ell (({v_L^I}^2)-({v_R^I}^2)) \in {\bf Z} \}.\eqno(4.20)$$
{}From the conditions (4.16), $\Gamma^{\prime D,D}$ is lorentzian even
self-dual lattice if $\Gamma^{D,D}$ is. Therefore, the total physical
Hilbert space of the ${\bf Z}_N$-orbifold model is equivalent to that of
the torus model associated with the lattice $\Gamma^{\prime D,D}$.

\vskip 10mm

\section{5. Asymmetric ${\bf Z}_N$-Orbifold Models}
\section{with Twist-Untwist Intertwining Currents}

In this section we present a list of asymmetric ${\bf Z}_N$-orbifold
models with
(1,0) states in twisted sectors and show that these states correspond to
twist-untwist intertwining currents which enlarge symmetries in each
untwisted or twisted sector to larger symmetries of the total Hilbert
space.

\vskip 5mm

\noindent
5.1. Asymmetric ${\bf Z}_2$-orbifold models with twist-untwist
intertwining currents

\vskip 2mm

Let us first consider asymmetric ${\bf Z}_2$-orbifolds. The lattice $
\Lambda$ in
eq. (3.1) is taken to be a root lattice of a simply-laced semi-simple Lie
algebra $G$ having the ${\bf Z}_2$-automorphism discussed in sect. 3. The
squared
length of the root vectors in the root lattice $\Lambda$ is normalized to
two. The ${\bf Z}_2$-transformation is defined by
$$
(X^I_L,X^I_R)\rightarrow(-X^I_L,X^I_R),\ (I=1,\dots,D).
\eqno(5.1)
$$
In this case, the necessary and sufficient conditions for modular
invariance
are
$$
D=0\ \ {\rm mod}\ 8,
\eqno(5.2)
$$
$$
2(p^I_L)^2=0\ \ {\rm mod}\ 2\ \ \ {\rm for\ all}\ p^I_R\in{\Gamma_0}^\ast,
\eqno(5.3)
$$
where
$$
\eqalign{\Gamma_0  &  =\{\ p^I_R\ \vert \ (p^I_L=0,p^I_R)\in\Gamma^{D,D}\
\}\cr
                   &  =\{\ p^I_R\ \vert \ p^I_R\in\Lambda\ \}.\cr
        }\eqno(5.4)
$$
The first condition (5.2) comes from eq. (2.10) because the conformal
weight
of the ground states in the twisted sector is given by (${D\over 16}$,0).
The condition (5.2) requires that the dimension $D$ of the root lattice
$\Lambda$ of a Lie algebra $G$ must be a multiple of eight, that is,
${\rm rank}G$
must be a multiple of eight.
For our purpose, it is sufficient to consider
models possessing (1,0) twisted states although some of these (1,0)
twisted
states might be unphysical. Thus it is sufficient to consider only the
cases
of $D=8$ and 16 since there appear no (1,0) states in the twisted sector
if
$D>16$. For $D=16$ the ground states in the twisted sector have the
conformal
weight (1,0). For $D=8$, the ground states in the twisted sector have the
conformal weight (${1\over 2}$,0) but the first excited states
have the conformal
weight (1,0) because the left-moving oscillators are expanded in
half-odd-integral modes in the twisted sector.

In sect. 3
we have obtained
the simply-laced Lie algebras
having the ${\bf Z}_2$-
automorphism.
Taking account of the condition (5.3),
we conclude that $G$ must be products of
$SO(8)$, $SO(16)$, $SO(24)$, $SO(32)$ and $E_8$ with
${\rm rank}G=8 \ {\rm or} \ 16$. All the models we have to investigate
are given in table 2.

In the following, we will mainly concentrate on the left-moving Hilbert
space. The $G_0$ in table 2 denotes the ${\bf Z}_2$-invariant subalgebra
of $G$,
which is the symmetry in each (untwisted or twisted) left-moving physical
Hilbert space. However, since there appear twisted states with the
conformal
weight (1,0), the symmetry $G_0$ will be \lq enlarged":
The physical (i.e., ${\bf Z}_2$-invariant) (1,0) states in the untwisted
sector
correspond to the adjoint representation of $G_0$. Each physical (1,0)
state
in the twisted sector corresponds to an intertwining current which
converts
untwisted states to twisted states, and vice versa. Thus the physical
(1,0)
states in the untwisted sector together with the physical (1,0) states in
the
twisted sector will form an adjoint representation of a larger group
$G^\prime$ than $G_0$, which is the full symmetry of the total physical
Hilbert space.

To see what the full symmetry $G^\prime$ is, we may rewrite each
asymmetric
${\bf Z}_2$-orbifold model into an equivalent torus model
using the \lq torus-orbifold equivalence", as explained in sect. 4. The
torus model which is equivalent to the asymmetric ${\bf Z}_2$-orbifold
model
with $\Gamma^{D,D}$ is specified by the following momentum lattice:
$$
\Gamma^{\prime D,D}=\{\ (p^{\prime I}_L,p^{\prime I}_R)
\in\bigcup^1_{\ell=0}[\Gamma^{D,D}+\ell(v^I_L,0)]
\ \vert\ p^\prime_L\cdot v_L-{1 \over 2}\ell(v_L^I)^2\in {\bf Z}\ \},
\eqno(5.5)
$$
where the shift vector $v^I_L$ for each model is given in an appendix. It
is
easy to determine the full symmetry $G^\prime$ from the lattice
${\Gamma^\prime}^{D,D}$.
The results are summarized in table 2.
It is interesting to note that $G^\prime$ is equal
to $G$ ${\rm for}\ D=8$
although $G^\prime$ is not necessarily the same as $G$.
In fact, it is not difficult to show that the momentum lattice
$\Gamma^{\prime D,D}$ is isomorphic to $\Gamma^{D,D}$ for each model with
$G^\prime=G$.

\vskip 5mm

\noindent
5.2. Asymmetric ${\bf Z}_N$-orbifold models with twist-untwist
intertwining currents

\vskip 2mm

Let us consider asymmetric ${\bf Z}_N$-orbifolds. The lattice $\Lambda$ in
eq.
(3.1) is taken to be a root lattice of a simply-laced semi-simple Lie
algebra $G$ having a ${\bf Z}_N$-automorphism discussed in sect. 3. The
squared
length of the root vectors in the root lattice $\Lambda$ is normalized
to two.

Let $g$ be a group element of the cyclic group ${\bf Z}_N$. The action of
$g$
on the string coordinate is defined by
$$
g:(X^I_L,X^I_R)\rightarrow (U^{IJ}X^J_L,X^I_R),\ (I,J=1,\dots,D),
\eqno(5.6)
$$
where $U$ is a $D\times D$ rotation matrix which satisfies $U^N={\bf 1}$
and
where $U$ and all its powers $U^\ell\ (\ell=1,\dots,N-1)$ are assumed to
have no fixed directions.

Let $N_\ell$ be the minimum positive integer such that
$(g^\ell)^{N_\ell}=1$. As discussed in sect. 2, $D$ must satisfy
$$
D=0\ \ {\rm mod}\ \varphi (N),
\eqno(5.7)
$$
and
$$
D=0\ \ {\rm mod}\ {\varphi(N_\ell)\over {N_\ell h_\ell^\prime}}
\ \ \ {\rm for}\ \ell=1,\dots,N-1,
\eqno(5.8)
$$
where
$$
h_\ell^\prime
={1\over4}
\sum_
{\scriptstyle (m,N_\ell)=1
\atop
\scriptstyle m=1,\dots,N_\ell-1}
{m\over N_\ell}\Bigl(1-{m\over N_\ell}\Bigr).
\eqno(5.9)
$$
Modular invariance further requires
$$
N_\ell(p^I_R)^2
=0\ \ {\rm mod}\ 2\ \ \ {\rm for\ all}\ p^I_R\in{\Gamma_0}^\ast ,
\eqno(5.10)
$$
where
$$
\eqalign{\Gamma_0  &  =\{\ p^I_R\ \vert \ (p^I_L=0,p^I_R)\in\Gamma^{D,D}\
\}\cr
                   &  =\{\ p^I_R\ \vert \ p^I_R\in\Lambda\ \}.\cr
        }\eqno(5.11)
$$
For our purpose, it is sufficient to consider models possessing (1,0)
twisted states although some of these (1,0) twisted states might be
unphysical. The conditions (5.7) and (5.8) and the existence of (1,0)
states in twisted sectors restrict the allowed dimensions $D$ for a given
order $N$. The results have been summarized in the fourth column of
table 1 up to $N=30$.

In sect. 3, we have classified the ${\bf Z}_N$-automorphism of the
simply-laced
Lie algebras and have shown that the allowed orders $N$ are restricted
to $N=2$,3,4,5,6,7,8,9,
10,11,12,13,15,16,17,19,20,23,24 and 30.
The conditions (5.7), (5.8) and (5.10) further restrict the
possible orders $N$ and lattices $\Lambda$.

All the models we have to investigate are given in tables 2$-$6. In
the following, we will mainly concentrate on the left-moving Hilbert
space. The $G_0$ in tables 2$-$6 denotes the ${\bf Z}_N$-invariant
subalgebra
of $G$, which is the symmetry in each (untwisted or twisted) left-moving
physical Hilbert space. However, since there appear twisted states with
the conformal weight (1,0), the symmetry $G_0$ will be \lq enlarged":
The physical (i.e., ${\bf Z}_N$-invariant) (1,0) states in the untwisted
sector
correspond to the adjoint representation of $G_0$. Each physical (1,0)
state in the twisted sector corresponds to an intertwining current which
converts untwisted states to twisted states, and vice versa. Thus, the
physical (1,0) states in the untwisted sector together with the physical
(1,0) states in the twisted sector will form an adjoint
representation of a larger group $G^\prime$ than $G_0$, which is the full
symmetry of the total physical Hilbert space.

To see what the full symmetry $G^\prime$ is, we may rewrite each
asymmetric
${\bf Z}_N$-orbifold model into an equivalent torus model using the
\lq torus-orbifold equivalence", as explained in sect. 4. The torus model
which is equivalent to the asymmetric ${\bf Z}_N$-orbifold model with
$\Gamma^{D,D}$ is specified by the following momentum lattice:
$$
\Gamma^{\prime D,D}=\{\ (p^{\prime I}_L,p^{\prime I}_R)
\in\bigcup^{N-1}_{\ell=0}[\Gamma^{D,D}+\ell(v^I_L,0)]
\ \vert \ p^{\prime}_L\cdot v_L-{1\over2}\ell(v_L^I)^2
\in {\bf Z} \ \},
\eqno(5.12)
$$
where the shift vector $v^I_L$ must satisfy
$$
\eqalignno{
Nv^I_L&\in\Lambda,
&(5.13)
\cr
\ell v^I_L&\not\in\Lambda\ (\ell=1,\dots,N-1).
&(5.14)
\cr
}$$
The shift vector $v^I_L$ can be chosen to satisfy
$$
{1\over2}(v^I_L)^2=h_1,
\eqno(5.15)
$$
where $h_1$ is the left-conformal weight of the ground states in the
$g$-twisted sector. For each model the shift vector $v^I_L$ which
satisfies
the conditions (5.13), (5.14) and (5.15) are explicitly given in an
appendix.
It is easy to determine the full symmetry $G^\prime$ from the lattice
$\Gamma^{\prime D,D}$.
The results are summarized in tables 2$-$6.
It is interesting to note that in some cases
$G^\prime$ is equal to $G$ although $G^\prime$ is not necessarily the same
as
$G$.
In fact, it is not difficult to show that the momentum lattice
$\Gamma^{\prime D,D}$ is isomorphic to $\Gamma^{D,D}$ for each model with
$G^\prime=G$.

We should make a comment on the case of the asymmetric ${\bf
Z}_{30}$-orbifold model
with $G=(E_8)^3$. In this model, there would appear (1,0) states in the
$g^\ell$-twisted sector with $\ell=2$,3,4,5,8,9,14,16,21,22,25,26,27 and
28.
However, all such states are not physical and are removed from the
physical
spectrum. This means that the lattice $\Gamma^{\prime 24,24}$ of the
equivalent torus model to the asymmetric ${\bf Z}_{30}$-orbifold model
contains no
vectors with $(p^{\prime I}_L)^2=2$ and ${p^\prime}^I_R=0$.
Since the root
lattice of $(E_8)^3$ is even self-dual, $\Gamma^{\prime 24,24}$ can be
written as
$$
\Gamma^{\prime 24,24}=\Gamma^{24,0}\oplus\Gamma^{0,24}_{(E_8)^3},
\eqno(5.16)
$$
where $\Gamma^{0,24}_{(E_8)^3}$ is the root lattice of $(E_8)^3$. Since
$\Gamma^{24,0}$ must be even self-dual and contains no vectors of norm
two,
it must be the Leech lattice [35].

\vskip 5mm

\noindent
5.3. The other choices of the squared length of the root vectors

\vskip 2mm

Let us consider asymmetric ${\bf Z}_N$-orbifold models, where the
squared length of the
root vectors in the root lattice $\Lambda$ of eq. (3.1) is normalized to
$2m\ (m=2,3,\dots)$. The ${\bf Z}_N$ -transformation is defined in eq.
(5.6). In
sect. 3 we have proved
that such consistent asymmetric orbifold models are
only the asymmetric ${\bf Z}_2$-orbifold models with $m=2$, $U=-{\bf 1}$
and
$G=(E_8)^\ell\ (\ell=1,2,\dots)$. Since we are interested in asymmetric
orbifold models with (1,0) twisted states, it is sufficient to consider
only the cases of $D=8\ {\rm and}\ 16$, i.e.,
$\Lambda=\sqrt2\Lambda_R(G)$ with
$G=E_8$ and $(E_8)^2$, where $\Lambda_R(G)$ denotes the root lattice
which is spanned by the root vectors with the squared length two.

Since there is no momentum such that $(p^I_L)^2=2$ and $p^I_R=0$ in the
untwisted sector, there are no (1,0) physical states in the untwisted
sector. In the twisted sector, there appear 8 and 1 (1,0) states for the
models with
$\Lambda=\sqrt2\Lambda_R(E_8)$ and $\sqrt2\Lambda_R((E_8)^2)$,
respectively
and they are found to be physical. Therefore, we may conclude
that the full symmetries of the total Hilbert spaces are
$(U(1))^8\ {\rm and}\ U(1)$ for
the models with
$\Lambda=\sqrt2\Lambda_R(E_8)\ {\rm and}\ \sqrt2\Lambda_R((E_8)^2)$,
respectively
although there is no symmetry
within each (untwisted or twisted) physical Hilbert space.

It should be noticed that the partition function of the above asymmetric
${\bf Z}_2$-orbifold model with $\Lambda=\sqrt2\Lambda_R(E_8)$ can be
shown to
be identical to that of the torus model
with which we have just started to
construct this orbifold model [17]. Hence this orbifold model can
probably be rewritten into the torus model although the
\lq torus-orbifold equivalence" discussed in sect. 4 is not applicable to
this case
{\footnote
{$^\star$}
{This orbifold model seems to give a counterexample of ref. [36].}}.

\vskip 10mm

\section{6. Supersymmetry Enhancement}

In this section we will consider $E_8\times E_8$ heterotic strings on
asymmetric ${\bf Z}_N$-orbifolds. The orbifold models are to be regarded
only as illustrative of the new features of asymmetric orbifolds and not
of a direct phenomenological relevance. In the light cone, right-movers
consist of eight bosons $X^i_R(i=1, \dots, 8)$ and eight
Neveu-Schwarz-Ramond (NSR) fermions $\lambda^i_R\ (i=1, \dots, 8)$. These
complete a right-moving superstring. Left-movers consist of eight bosons
$X^i_L\ (i=1, \dots, 8)$ and another sixteen bosons $\phi^I_L\ (I=1,
\dots, 16)$.

     A four-dimensional string can be constructed by compactifying the 22
left-moving $(X^a_L, \phi^I_L; a=3, \dots, 8, \ I=1, \dots, 16)$ and 6
right-moving $(X^a_R; a=3, \dots, 8)$ extra bosonic coordinates on a
torus. Then momenta $(p^a_L, P^I_L; p^a_R)$ of the string coordinates $
(X^a_L, \phi^I_L; X^a_R)$ lie on a (22+6)-dimensional lorentzian even
self-dual lattice $\Gamma^{22,6}$, i.e.,
$$(p^a_L, P^I_L; p^a_R) \in \Gamma^{22,6} \quad (a=3, \dots, 8, \ I=1,
\dots, 16). \eqno(6.1)$$
The lattice $\Gamma^{22, 6}$ is assumed to be of the form
$$ \Gamma^{22,6}=\Gamma^{6,6}\oplus \Gamma^{16,0}_{E_8 \times E_8}, \eqno
(6.2)$$
where $\Gamma^{16,0}_{E_8 \times E_8}$ is the root lattice of $E_8 \times
E_8$ on which $P^I_L\ (I=1, \dots , 16)$ lies. The (6+6)-dimensional
lattice $\Gamma^{6,6}$ has to be even self-dual and is taken to be the
lattice defined in (3.1). The lattice $\Lambda$ in eq. (3.1) is a root
lattice of a simply-laced semi-simple Lie algebra with rank $G = 6$. The
squared length of the root vectors is normalized to two.

   To construct an asymmetric ${\bf Z}_N$-orbifold, we will consider the
following ${\bf Z}_N$-transformation:
$$ \eqalign{g:X^a_L& \rightarrow X^a_L, \cr
              X^a_R& \rightarrow (UX_R)^a\quad , \quad (a=3, \dots , 8),
\cr} \eqno(6.3)$$
where $U$ is a 6-dimensional rotation matrix with $U^N={\bf 1}$. The ${\bf
Z}_N$-transformation must be an automorphism of the lattice $
\Gamma^{6,6}$, i.e.,
$$(p^a_L, (Up_R)^a) \in \Gamma^{6,6} \quad {\rm for\ all}\ (p^a_L, p^a_R)
\in \Gamma^{6,6}. \eqno(6.4) $$
The action of $g$ on the NSR fermions is the same as the action on $X^a_R$
to preserve world sheet supersymmetry [2]. The action of $g$ on the
remaining fields is taken to be trivial.

   As before, we will assume that $U$ and all its powers $U^\ell \ (\ell
=1, \dots , N-1)$ have no fixed direction. Then it is not difficult to
show that there are only four modular invariant asymmetric ${\bf
Z}_N$-orbifold models: two ${\bf Z}_3$-orbifold models with $G=(SU(3))^3$
and $E_6$, one ${\bf Z}_7$-orbifold model with $G=SU(7)$ and one ${\bf
Z}_9$-orbifold model with $G=E_6$. Let ${\rm exp}(i2\pi v^t)$ and ${\rm
exp}(-i2\pi v^t)\ (t=1,2,3)$ be the eigenvalues of $U$. The $v^t$ will be
taken to be
$$\eqalignno{
v^t&=({1\over 3}, {1\over 3}, {2\over 3})\quad {\rm for\ the\ }{\bf Z}_3{
\rm -models, }\cr
   &=({1\over 7}, {2\over 7}, {3\over 7})\quad {\rm for\ the\ }{\bf Z}_7{
\rm -model, }\cr
   &=({1\over 9}, {2\over 9}, {5\over 9})\quad {\rm for\ the\ }{\bf Z}_9{
\rm -model. }&(6.5)\cr
}$$
It should be noted that the ${\bf Z}_3$-and ${\bf Z}_7$-transformations
leave one unbroken space-time supersymmetry while the ${\bf
Z}_9$-transformation leaves no unbroken space-time supersymmetry [1]. That
is, N=1\ (N=0) space-time supersymmetry survives in each of untwisted and
twisted sectors for the ${\bf Z}_3((SU(3))^3)$-, ${\bf Z}_3(E_6)$- and ${
\bf Z}_7(SU(7))$- $\bigl( {\bf Z}_9(E_6)-\bigr) $ orbifold models. In
fact, for the ${\bf Z}_3((SU(3))^3)$-, ${\bf Z}_3(E_6)$- and ${\bf Z}_7(SU
(7))$- $\bigl( {\bf Z}_9(E_6)-\bigr) $ orbifold models an N=1\ (N=0)
supergravity multiplet coupled to an N=1\ (N=0) super Yang-Mills multiplet
with $E_8\times E_8\times G$ with $G=(SU(3))^3, E_6$ and $SU(7)\ (E_6)$
appears in the untwisted sector, respectively {\footnotes{$^\star$}{The N
=0 super multiplet means the N=1 super multiplet without the fermion
contents.}}.

   This is not the end of the story. A number of massless states appear in
the twisted sectors. The degeneracy $d_\ell$ of the ground states in the
$g^\ell$-sector $(\ell=1, \dots , N-1)$ is given by
$$\eqalignno{
d_\ell&=1 \qquad \ell=1,2\quad {\rm for\ the\ }{\bf Z}_3((SU(3))^3){\rm
-model}, \cr
d_\ell&=3 \qquad \ell=1,2\quad {\rm for\ the\ }{\bf Z}_3(E_6){\rm -model},
\cr
d_\ell&=1 \qquad \ell=1, \dots ,6\quad {\rm for\ the\ }{\bf Z}_7(SU(7)){
\rm -model}, \cr
d_\ell&=
          \cases{1 \quad \ell&=1,2,4,5,7,8\quad {\rm for\ the\ ${\bf Z}_9
(E_6)$-model}. \cr
        3 \quad \ell&=3,6 \cr}
&(6.6)\cr
}$$
For the ${\bf Z}_3((SU(3)^3)$-orbifold model, there appear one massless
spin $3\over 2$ fermion, two $U(1)$ gauge bosons, one massless spin $1
\over 2$ fermion and two massless scalars in the twisted sectors. The
massless spin $1\over 2$ fermion and the two massless scalars belong to
the adjoint representation of $E_8\times E_8\times (SU(3))^3$. Those
massless fields in the twisted sectors together with the massless fields
in the untwisted sector will form an N=2 supergravity multiplet coupled to
an N=2 super Yang-Mills multiplet with $E_8\times E_8\times (SU(3))^3$.

   For the ${\bf Z}_3(E_6)$-and ${\bf Z}_7(SU(7))$-orbifold models, there
appear three massless spin $3\over 2$ fermions, six $U(1)$ gauge bosons,
three massless spin $1\over 2$ fermions and six massless scalars in the
twisted sectors. The three massless spin $1\over 2$ fermions and the six
massless scalars belong to the adjoint representation of $E_8\times E_8
\times G$ with $G=E_6$ or $SU(7)$. Thus those massless fields in the
twisted sectors together with the massless fields in the untwisted sector
will form an N=4 supergravity multiplet coupled to an N=4 super Yang-Mills
multiplet with $E_8\times E_8\times E_6$ or $E_8\times E_8\times SU(7)$.

   For the ${\bf Z}_9(E_6)$-orbifold model, there appear three massless
spin $3\over 2$ fermions, six $U(1)$ gauge bosons, three massless spin $1
\over 2$ fermions and six massless scalars in the $g^\ell$-twisted sectors
with $\ell=1,2,4,5,7$ and 8. All these massless states are found to be
physical. The three massless spin $1\over 2$ fermions and the six massless
scalars belong to the adjoint representation of $E_8\times E_8\times E_6$.
There would also appear three massless spin $3\over 2$ fermions, six $U
(1)$ gauge bosons, three massless spin $1\over 2$ fermions and six
massless scalars in the $g^\ell$-twisted sectors with $\ell=3$ and 6.
However, these massless fields are not always physical. In fact, a
detailed analysis of the one loop vacuum amplitude tells us that only one
massless spin $3\over 2$ fermion and one massless spin $1\over 2$ fermion
survive as physical states. Other states, in particular, all bosonic
states are removed from the physical spectrum.
Therefore, the massless fields in the twisted sectors together with the
massless fields in the untwisted sector will form an N=4 supergravity
multiplet coupled to an N=4 super Yang-Mills multiplet with $E_8\times E_8
\times E_6$ although no supersymmetry survives in the untwisted sector.

   We have investigated a restricted class of heterotic strings on
asymmetric orbifolds and have found that N=1, 1, 1 and 0 space-time
supersymmetry in each of the untwisted and twisted sectors is enlarged to
N=2, 4, 4 and 4 space-time supersymmetry in the total Hilbert spaces for
the ${\bf Z}_3((SU(3))^3)$-, ${\bf Z}_3(E_6)$-, ${\bf Z}_7(SU(7))$- and ${
\bf Z}_9(E_6)$-orbifold models, respectively although this conclusion has
not rigorously been proved. It is worth while noting that space-time
supersymmetry \lq enhancement" discussed above will also imply
{\it world-sheet}
supersymmetry \lq enhancement"[37].

\vskip 10mm

\section{7. Discussions}

We have presented a list of asymmetric ${\bf Z}_N$-orbifold models to
possess
(1,0) twisted states. We have found that these twisted states play a role
of intertwining currents which convert untwisted string states to twisted
ones, and vice versa and that full symmetries of such orbifold models are
larger than symmetries in each of the untwisted and twisted sectors
{\footnote
{$^\star$}
{The asymmetric ${\bf Z}_{30}$-orbifold model with $(E_8)^3$ is an
exception because
all (1,0) twisted states are unphysical.}}.
As mentioned in the introduction, this symmetry \lq enhancement" is
inherent
in asymmetric orbifolds.
We have seen that the conditions for the momentum
lattices with ${\bf Z}_N$-automorphisms, modular invariance and the
existence of
(1,0) twisted states put severe restrictions on such orbifold models. It
may be interesting to point out that for all the orbifold models which we
encountered in this paper (1,0) twisted states can appear only for
$D\leq24$. The maximum dimension $D=24$ is equal to the transverse
dimension
of the bosonic string theory in the light-cone gauge.

In tables 2$-$6, $G_0$ denotes
the symmetries of the untwisted and twisted
sectors and $G^\prime$ denotes the full symmetries of the total Hilbert
spaces. Each orbifold model listed in tables 2$-$6 has a symmetry
$G^\prime\times G$, where the symmetry $G^\prime\ (G)$ comes from the
left- (right-) moving modes. In general, the symmetry $G^\prime$ of the
left-movers is different from the symmetry $G$ of the right-movers due to
asymmetric nature of the orbifolds. For some orbifold models, the symmetry
$G^\prime$ is, however, equal to $G$ and hence asymmetric nature
disappears. Specifically, for $D\leq10$, all the orbifold models we
considered have this property.

All the orbifold models in tables 2$-$6 have been shown
to be equivalently
rewritten into the torus models because the ${\bf Z}_N$-transformations
are inner
automorphisms of the momentum lattices $\Gamma^{D,D}$. We may
schematically
write the \lq torus-orbifold equivalence" as $T^D/P\simeq T^{\prime D}$,
where
$T^D/P\ \ (T^{\prime D})$ denotes a $D$-dimensional orbifold (torus)
model. For
symmetric orbifold models, there exist no intertwining currents and hence
$G^\prime$ must be equal to $G_0$. This implies that $T^D\not=T^{\prime
D}$
for symmetric orbifolds. On the other hand, for asymmetric orbifold
models,
there exist examples of $T^D/P\simeq T^{\prime D}$ with
$T^D=T^{\prime D}$. Specifically, for $D\leq10$, all the orbifold models
we considered have
this property. It would be of interest to examine the above
peculiarity for lower dimensional asymmetric orbifolds. More interesting
observation is given in subsect. 5.3. for $D=8$. Since rank of $G_0$ is
less
than rank of $G\ ({\rm rank}G_0=0\ {\rm and}\ {\rm rank}G=8)$,
the ${\bf Z}_2$-transformation will
correspond to an outer automorphism of the momentum lattice. Hence it
seems
that this orbifold model could not be rewritten into a torus model.
However,
we have shown that $G^\prime=G=(U(1))^8$ and can prove that the one loop
partition function of the orbifold model
is identical to that of the torus
model which we just started with to define this orbifold model.
This result strongly
suggests that the orbifold model is equivalent to the torus model even
though
the ${\bf Z}_2$-transformation is an outer automorphism and that the
\lq torus-orbifold equivalence" stated in sect. 4 should be replaced by
the
following statement: Any closed bosonic string theory compactified on an
orbifold is equivalent to that on a torus if rank of the symmetry of the
{\it total} physical Hilbert space is equal
to the dimension of the orbifold.

In sect. 6
we have considered the $E_8\times E_8$ heterotic strings on the
asymmetric ${\bf Z}_N$-orbifolds and found supersymmetry \lq enhancement",
instead of
gauge symmetry \lq enhancement". This mechanism will lead to a new class
of
four-dimensional string models with N=1 space-time supersymmetry:
There is
no unbroken supersymmetry in each of untwisted and twisted sectors but
there
exists a space-time supercharge which convert untwisted string states to
twisted string states with opposite statistics, and vice versa.
It would be of
importance to study the new class of four-dimensional string models with
N=1 space-time supersymmetry.

\vskip 10mm

\acknowledgments

One of us (M.S) would like to thank J.L.Petersen and K.Ito for useful
discussions and also would like to acknowledge the hospitality for
the Niels Bohr Institute where part of this work was done.

\endpage

\section{Appendix}

In this appendix, we will give the shift vectors $v_L^I$ which are
introduced in rewriting the asymmetric ${\bf Z}_N$-orbifold models into
the equivalent torus models. It will be sufficient to give the
shift vectors only for the simple Lie algebra root lattices.

In the usual orthonormal basis, the root lattices $\Lambda_R$
of the simple Lie algebras are given by

\noindent \hskip 17mm
$
\Lambda_R(SU(n+1))=\{\ (m_1,m_2,\dots,m_{n+1})\ \vert\ m_i\in{\bf Z},\
\sum_{i=1}^{n+1}m_i=0\ \},
$

\noindent \hskip 17mm
$
\Lambda_R(SO(2n))=\{\ (m_1,m_2,\dots,m_n)\ \vert\ m_i\in{\bf Z},\ \sum_{i
=1}^nm_i\in2{\bf Z}\ \},
$

\noindent \hskip 17mm
$
\Lambda_R(E_8)=\Lambda_R(SO(16))\cup[\ \Lambda_R(SO(16))+({1\over2},{1
\over2},{1\over2},{1\over2},{1\over2},{1\over2},{1\over2},{1\over2})\ ],
$

\noindent \hskip 17mm
$
\Lambda_R(E_6)=\{\ p^I\in\Lambda_R(E_8)\ \vert\ p\cdot(e_6+e_8)=p\cdot(e_7
+e_8)=0\ \},
$

\noindent
where we have normalized the squared length of the root vectors to two.

The shift vectors $v_L^I$ for the asymmetric ${\bf Z}_N$-orbifolds
which satisfy the conditions (5.13), (5.14) and (5.15) are given
in the usual orthonormal basis as follows:

(a) Asymmetric ${\bf Z}_2$-orbifolds

\noindent \hskip 17mm $v_L^I(E_8)={1\over2}(2,0,0,0,0,0,0,0),$

\noindent \hskip 17mm $v_L^I(SO(32))={1\over2}
(1,1,1,1,1,1,1,1,0,0,0,0,0,0,0,0),$

\noindent \hskip 17mm $v_L^I(SO(24))={1\over2}(1,1,1,1,1,1,0,0,0,0,0,0),$

\noindent \hskip 17mm $v_L^I(SO(16))={1\over2}(1,1,1,1,0,0,0,0),$

\noindent \hskip 17mm $v_L^I(SO(8))={1\over2}(1,1,0,0),$

(b) Asymmetric ${\bf Z}_3$-orbifolds

\noindent \hskip 17mm $v_L^I(E_8)={1\over3}(0,2,1,0,-1,1,0,1),$

\noindent \hskip 17mm $v_L^I(E_6)={1\over3}(0,1,2,0,1,0,0,0),$

\noindent \hskip 17mm $v_L^I(SU(3))={1\over3}(1,0,-1),$

(c) Asymmetric ${\bf Z}_4$-orbifold

\noindent \hskip 17mm $v_L^I(E_8)={1\over4}(0,-1,2,1,0,-1,2,1),$

\noindent \hskip 17mm $v_L^I(SO(32))={1\over4}
(0,-1,2,1,0,-1,2,1,0,-1,2,1,0,-1,2,1),$

\noindent \hskip 17mm $v_L^I(SO(24))={1\over4}
(0,-1,2,1,0,-1,2,1,0,-1,2,1),$

\noindent \hskip 17mm $v_L^I(SO(16))={1\over4}(0,-1,2,1,0,-1,2,1),$

\noindent \hskip 17mm $v_L^I(SO(8))={1\over4}(0,-1,2,1),$

(d) Asymmetric ${\bf Z}_5$-orbifold

\noindent \hskip 17mm $v_L^I(E_8)={1\over5}(0,-1,3,2,1,0,-1,2),$

\noindent \hskip 17mm $v_L^I(SU(5))={1\over5}(2,1,0,-1,-2),$

(e) Asymmetric ${\bf Z}_6$-orbifold

\noindent \hskip 17mm $v_L^I(E_8)={1\over6}(0,-1,-2,3,2,1,0,1),$

(f) Asymmetric ${\bf Z}_7$-orbifolds

\noindent \hskip 17mm $v_L^I(SU(7))={1\over7}(3,2,1,0,-1,-2,-3),$

(g) Asymmetric ${\bf Z}_9$-orbifold

\noindent \hskip 17mm $v_L^I(E_6)={1\over9}(0,-1,-2,-3,-4,-2,-2,2),$

(h) Asymmetric ${\bf Z}_{10}$-orbifolds

\noindent \hskip 17mm $v_L^I(E_8)={1\over10}(0,-3,4,1,-2,5,2,1),$

(i) Asymmetric ${\bf Z}_{11}$-orbifolds

\noindent \hskip 17mm $v_L^I(SU(11))={1\over11}
(5,4,3,2,1,0,-1,-2,-3,-4,-5),$

(j) Asymmetric ${\bf Z}_{13}$-orbifolds

\noindent \hskip 17mm $v_L^I(SU(13))={1\over13}
(6,5,4,3,2,1,0,-1,-2,-3,-4,-5,-6),$

(k) Asymmetric ${\bf Z}_{15}$-orbifolds

\noindent \hskip 17mm $v_L^I(E_8)={1\over15}(0,-1,-2,-3,-4,-5,-6,7),$

(l) Asymmetric ${\bf Z}_{17}$-orbifold

\noindent \hskip 17mm $v_L^I(SU(17))={1\over17}
(8,7,6,5,4,3,2,1,0,-1,-2,-3,-4,-5,-6,-7,-8),$

(m) Asymmetric ${\bf Z}_{19}$-orbifold

\noindent \hskip 17mm $v_L^I(SU(19))$

\noindent \hskip 17mm $={1\over19}
(9,8,7,6,5,4,3,2,1,0,-1,-2,-3,-4,-5,-6,-7,-8,-9),$

(n) Asymmetric ${\bf Z}_{20}$-orbifold

\noindent \hskip 17mm $v_L^I(E_8)={1\over20}(0,1,1,2,2,3,4,15),$

(o) Asymmetric ${\bf Z}_{23}$-orbifold

\noindent \hskip 17mm $v_L^I(SU(23))$

$={1\over23}(11,10,9,8,7,6,5,4,3,2,1,0,
                   -1,-2,-3,-4,-5,-6,-7,-8,-9,-10,-11),$

(p) Asymmetric ${\bf Z}_{30}$-orbifold

\noindent \hskip 17mm $v_L^I(E_8)={1\over30}(0,-1,-2,-3,-4,-5,-6,-23).$

\endpage

\references

\item{[1]} L. Dixon, J.A. Harvey, C. Vafa and E. Witten, Nucl. Phys. {\bf
B261} (1985) 678; {\bf B274} (1986) 285.
\item{[2]} H. Kawai, D. Lewellen and A.H. Tye, Phys. Rev. Lett. {\bf 57}
(1986) 1832; Nucl. Phys. {\bf B288} (1987) 1;
\item{   } I. Antoniadis, C. Bachas and C. Kounnas, Nucl. Phys. {\bf B289}
(1987) 87.
\item{[3]} W. Lerche, A.N. Schellenkens and N.P. Warner, Phys. Rep. {\bf
177} (1989) 1.
\item{[4]} D. Gepner, Phys. Lett. {\bf B199} (1987) 380; Nucl. Phys. {\bf
B296} (1987) 757.
\item{[5]} Y. Kazama and H. Suzuki, Nucl. Phys. {\bf B321} (1989) 232.
\item{[6]} C. Vafa and N.P. Warner, Phys. Lett. {\bf B218} (1989) 51;
\item{    } W. Lerche, C. Vafa and N.P. Warner, Nucl. Phys. {\bf B324}
(1989) 427;
\item{    } P.S. Howe and P.C. West, Phys. Lett. {\bf B223} (1989) 377; {
\bf B244} (1989) 270.
\item{[7]} E.S. Fradkin and A.A. Tseytlin, Phys. Lett. {\bf B158} (1985)
316; Nucl. Phys. {\bf B261} (1985) 1;
\item{    } C.G. Callan, D. Friedan, E.J. Martinec and M.J. Perry, Nucl.
Phys. {\bf B262} (1985) 593;
\item{    } C.G. Callan, I.R. Klebanov and M.J. Perry, Nucl. Phys. {\bf
B278} (1986) 78;
\item{    } T. Banks, D. Nemeschansky and A. Sen, Nucl. Phys. {\bf B277}
(1986) 67.
\item{[8]} P. Candelas, G. Horowitz, A. Strominger and E. Witten, Nucl.
Phys. {\bf B258} (1985) 46.
\item{[9]} D.J. Gross, J.A. Harvey, E. Martinec and R. Rohm, Nucl. Phys. {
\bf B256} (1985) 253; {\bf B267} (1986) 75.
\item{[10]} K.S. Narain, M.H. Sarmadi and C. Vafa, Nucl. Phys. {\bf B288}
(1987) 551; {\bf B356} (1991) 163.
\item{[11]} A. Font, L.E. Ib\'a\~ nez, F. Quevedo and A. Sierra, Nucl.
Phys. {\bf B331} (1990) 421.
\item{[12]} J.A. Casas and C. Mu\~ noz, Nucl. Phys. {\bf B332} (1990) 189.
\item{[13]} Y. Katsuki, Y. Kawamura, T. Kobayashi, N. Ohtsubo, Y. Ono and
K. Tanioka, Nucl. Phys. {\bf B341} (1990) 611.
\item{[14]} A. Fujitsu, T. Kitazoe, M. Tabuse and H. Nishimura, Intern. J.
Mod. Phys. {\bf A5} (1990) 1529.
\item{[15]} E. Corrigan and T.J. Hollowood, Nucl. Phys. {\bf B304} (1988)
77.
\item{[16]} L. Dolan, P. Goddard and P. Montague, Nucl. Phys. {\bf B338}
(1990) 529.
\item{[17]} Y. Imamura, M. Sakamoto and M. Tabuse, Phys. Lett. {\bf B266}
(1991) 307.
\item{[18]} Y. Imamura, M. Sakamoto, T. Sasada and M. Tabuse, Kobe
preprint, KOBE-92-01 (1992).
\item{[19]} I. Frenkel, J. Lepowsky and A. Meurman, Proc. Conf. on Vertex
operators in mathematics and physics, ed. J. Lepowsky et al. (Springer
1984).
\item{[20]} R. Dijkgraaf, E. Verlinde and H. Verlinde, Commun. Math. Phys.
{\bf 115} (1988) 649;
\item{    } P. Ginsparg, Nucl. Phys. {\bf B295} (1988) 153.
\item{[21]} M. Sakamoto, Mod. Phys. Lett. {\bf A5} (1990) 1131; Prog.
Theor. Phys. {\bf 84} (1990) 351.
\item{[22]} K.S. Narain, Phys. Lett. {\bf B169} (1986) 41;
\item{    } K.S. Narain, M.H. Sarmadi and E. Witten, Nucl. Phys. {\bf
B279} (1987) 369.
\item{[23]} C. Vafa, Nucl. Phys. {\bf B273} (1986) 592.
\item{[24]} M. Sakamoto and M. Tabuse, Kobe preprint, KOBE-92-02 (1992).
\item{[25]} T. Horiguchi, M. Sakamoto and M. Tabuse, Kobe preprint,
KOBE-92-03 (1992).
\item{[26]} K. Ireland and M. Rosen, A classical introduction to modern
number theory, 2nd ed. (Springer-Verlag, 1990).
\item{[27]} F. Englert and A. Neveu, Phys. Lett. {\bf B163} (1985) 349.
\item{[28]} J.E. Humphreys, \ Introduction \ to \ Lie \ Algebras \ and \
Representation \ Theory \  (Springer, 1972).
\item{[29]} R.G. Myhill, Durham preprint, DTP-86/19 (1986).
\item{[30]} N. Bourbaki, {\sl \'El\'ements  de  Math\'ematique, Groupes et
Algebras  Lie }  (Hermann, Paris, 1960).
\item{[31]} E. Corrigan and D. Olive, Nuovo Cimento {\bf 11A} (1972) 749.
\item{[32]} Y. Kazama\quad  and\quad  H. Suzuki,\quad  Phys. Lett. {\bf
B192} (1987) 351;\quad  KEK preprint KEK-TH-165 (1987).
\item{[33]} B. Gato, Nucl. Phys. {\bf B322} (1989) 555.
\item{[34]} I. Frenkel and V. Ka\v c, Invent. Math. {\bf 62} (1980) 23;
\item{    } G. Segal, Commun. Math. Phys. {\bf 80} (1981) 301;
\item{    } P. Goddard and D. Olive, Intern. J. Mod. Phys. {\bf A1} (1986)
303.
\item{[35]} J. Leech, Can. J. Math. 19 (1967) 251.
\item{[36]} B. Rostand, Phys. Lett. {\bf B248} (1990) 89.
\item{[37]} T.Banks, L.J.Dixon, D.Friedan and E.Martinec, Nucl. Phys.
{\bf B299} (1988) 613;
\item{    } T.Banks and L.J.Dixon, Nucl. Phys. {\bf B307} (1988) 93.

\endpage

\nopagenumbers

\table
\noindent
Table 1. $\varphi(N)$ denotes the Euler function. In the third column allowed
values of $D$, which are consistent with the constraints (2.16) and (2.19),
are given. In the fourth column the dimensions $D$ in which (1,0) states
might appear in twisted sectors are given.

\vskip 2mm

\noindent
Table 2. $G_0$ denotes the ${\bf Z}_2$-invariant subgroup of $G$, which is
the symmetry in each sector and $G'$ denotes the full symmetry of the
total Hilbert space.

\vskip 2mm

\noindent
Table 3. $G_0$ denotes the ${\bf Z}_3$-invariant subgroup of $G$, which is
the symmetry in each sector and $G'$ denotes the full symmetry of the
total Hilbert space.

\vskip 2mm

\noindent
Table 4. $G_0$ denotes the ${\bf Z}_4$-invariant subgroup of $G$, which is
the symmetry in each sector and $G'$ denotes the full symmetry of the
total Hilbert space.

\vskip 2mm

\noindent
Table 5. $G_0$ denotes the ${\bf Z}_5$-invariant subgroup of $G$, which is
the symmetry in each sector and $G'$ denotes the full symmetry of the
total Hilbert space.

\vskip 2mm

\noindent
Table 6. $G_0$ denotes the ${\bf Z}_N$-invariant subgroup of $G$, which is
the symmetry in each sector and $G'$ denotes the full symmetry of the
total Hilbert space.

\endpage

\centerline{Table 1.}
\vskip 4mm

\centerline{
\vbox{
\offinterlineskip
\halign{\vrule#&\strut\ \hfil#\ \hfil&\vrule#&\hfil\quad#\hfil\quad&
\vrule#&\hfil\quad#\hfil\quad&\vrule#&\hfil\quad#\hfil\quad&\vrule#\cr
\noalign{\hrule}
height3pt&\omit&&\omit&&\omit&&\omit&\cr
& ${\bf Z}_N$  && $\varphi (N)$          && $D$        && $D$
&\cr
height3pt&\omit&&\omit&&\omit&&\omit&\cr
\noalign{\hrule}
height3pt&\omit&&\omit&&\omit&&\omit&\cr
& ${\bf Z}_2$  && 1    &&   8{\bf Z}     &&   8, 16        &\cr
height3pt&\omit&&\omit&&\omit&&\omit&\cr
& ${\bf Z}_3$  && 2    &&   6{\bf Z}     &&   6, 12, 18        &\cr
height3pt&\omit&&\omit&&\omit&&\omit&\cr
& ${\bf Z}_4$  && 2    &&   16{\bf Z}     &&   16        &\cr
height3pt&\omit&&\omit&&\omit&&\omit&\cr
& ${\bf Z}_5$  && 4    &&   4{\bf Z}     &&   4, 8, 12, 16, 20        &\cr
height3pt&\omit&&\omit&&\omit&&\omit&\cr
& ${\bf Z}_6$  && 2    &&   24{\bf Z}     &&   24        &\cr
height3pt&\omit&&\omit&&\omit&&\omit&\cr
& ${\bf Z}_7$  && 6    &&   6{\bf Z}     &&   6, 12, 18        &\cr
height3pt&\omit&&\omit&&\omit&&\omit&\cr
& ${\bf Z}_8$  && 4    &&   32{\bf Z}     &&  {\rm non}        &\cr
height3pt&\omit&&\omit&&\omit&&\omit&\cr
& ${\bf Z}_9$  && 6   &&   18{\bf Z}     &&   18        &\cr
height3pt&\omit&&\omit&&\omit&&\omit&\cr
& ${\bf Z}_{10}$  && 4    &&  8{\bf Z}     &&   8, 16, 24        &\cr
height3pt&\omit&&\omit&&\omit&&\omit&\cr
& ${\bf Z}_{11}$  && 10    &&   10{\bf Z}     &&   10, 20        &\cr
height3pt&\omit&&\omit&&\omit&&\omit&\cr
& ${\bf Z}_{12}$  && 4    &&   48{\bf Z}     &&   {\rm non}        &\cr
height3pt&\omit&&\omit&&\omit&&\omit&\cr
& ${\bf Z}_{13}$  && 12    &&   12{\bf Z}     &&   12        &\cr
height3pt&\omit&&\omit&&\omit&&\omit&\cr
& ${\bf Z}_{14}$  && 6    &&   24{\bf Z}     &&   24        &\cr
height3pt&\omit&&\omit&&\omit&&\omit&\cr
& ${\bf Z}_{15}$  && 8    &&   24{\bf Z}     &&   24        &\cr
height3pt&\omit&&\omit&&\omit&&\omit&\cr
& ${\bf Z}_{16}$  && 8    &&   64{\bf Z}     &&   {\rm non}        &\cr
height3pt&\omit&&\omit&&\omit&&\omit&\cr
& ${\bf Z}_{17}$  && 16    &&   16{\bf Z}     &&   16        &\cr
height3pt&\omit&&\omit&&\omit&&\omit&\cr
& ${\bf Z}_{18}$  && 6    &&   72{\bf Z}     &&   {\rm non}        &\cr
height3pt&\omit&&\omit&&\omit&&\omit&\cr
& ${\bf Z}_{19}$  && 18    &&   18{\bf Z}     &&   18        &\cr
height3pt&\omit&&\omit&&\omit&&\omit&\cr
& ${\bf Z}_{20}$  && 8    &&   16{\bf Z}     &&    16        &\cr
height3pt&\omit&&\omit&&\omit&&\omit&\cr
& ${\bf Z}_{21}$  && 12    &&   12{\bf Z}     &&   12, 24        &\cr
height3pt&\omit&&\omit&&\omit&&\omit&\cr
& ${\bf Z}_{22}$  && 10    &&   40{\bf Z}     &&   {\rm non}       &\cr
height3pt&\omit&&\omit&&\omit&&\omit&\cr
& ${\bf Z}_{23}$  && 22    &&   22{\bf Z}     &&   22        &\cr
height3pt&\omit&&\omit&&\omit&&\omit&\cr
& ${\bf Z}_{24}$  && 8    &&   96{\bf Z}     &&     {\rm non}        &\cr
height3pt&\omit&&\omit&&\omit&&\omit&\cr
& ${\bf Z}_{25}$  && 20    &&   20{\bf Z}     &&   20        &\cr
height3pt&\omit&&\omit&&\omit&&\omit&\cr
& ${\bf Z}_{26}$  && 12    &&   24{\bf Z}     &&   24        &\cr
height3pt&\omit&&\omit&&\omit&&\omit&\cr
& ${\bf Z}_{27}$  && 18    &&   54{\bf Z}     &&   {\rm non}       &\cr
height3pt&\omit&&\omit&&\omit&&\omit&\cr
& ${\bf Z}_{28}$  && 12    &&   48{\bf Z}     &&   {\rm non}        &\cr
height3pt&\omit&&\omit&&\omit&&\omit&\cr
& ${\bf Z}_{29}$  && 28    &&   28{\bf Z}     &&   {\rm non}       &\cr
height3pt&\omit&&\omit&&\omit&&\omit&\cr
& ${\bf Z}_{30}$  && 8    &&   24{\bf Z}     &&    24        &\cr
height3pt&\omit&&\omit&&\omit&&\omit&\cr
\noalign{\hrule}
}}}

\endpage

\noindent
\centerline{Table 2.}
\centerline{Asymmetric ${\bf Z}_2$-orbifold models with (1,0) twisted
states.}
\vskip 4mm

\centerline{
\vbox{\offinterlineskip
\halign{&\vrule#&\strut\quad\hfil#\hfil\quad&\vrule#&\quad\hfil#\hfil
\quad&\vrule#&\quad\hfil#\hfil\quad&\vrule# \cr
\noalign{\hrule}
height3pt&\omit      &&\omit             &&\omit            &\cr
&  $G$                && $G_0$            && $G'$             &\cr
height3pt&\omit      &&\omit             &&\omit            &\cr
\noalign{\hrule}
height6pt&\omit      &&\omit             &&\omit            &\cr
& $D=8$\qquad\qquad\qquad\quad &&                &&                &\cr
height3pt&\omit      &&\omit             &&\omit            &\cr
& $E_8$              && $SO(16)$         && $E_8$            &\cr
height3pt&\omit      &&\omit             &&\omit            &\cr
& $SO(16)$           && $(SO(8))^2$      && $SO(16)$         &\cr
height3pt&\omit      &&\omit             &&\omit            &\cr
& $(SO(8))^2$        && $(SU(2))^8$      && $(SO(8))^2$      &\cr
height6pt&\omit      &&\omit             &&\omit            &\cr
& $D=16$\qquad\qquad\qquad\  &&                &&                &\cr
height3pt&\omit      &&\omit             &&\omit            &\cr
& $(E_8)^2$          && $(SO(16))^2$     && $SO(32)$   &\cr
height3pt&\omit      &&\omit             &&\omit            &\cr
& $SO(32)$           && $(SO(16))^2$     && $E_8\times SO(16)$ &\cr
height3pt&\omit      &&\omit             &&\omit            &\cr
& $SO(24)\times SO(8)$ && $(SO(12))^2\times (SU(2))^4$ && $E_7\times SO
(12)\times (SU(2))^3$ &\cr
height3pt&\omit      &&\omit             &&\omit            &\cr
& $E_8\times SO(16)$ && $SO(16)\times (SO(8))^2$ && $SO(24)\times SO(8)$ &
\cr
height3pt&\omit      &&\omit             &&\omit            &\cr
& $(SO(16))^2$ && $(SO(8))^4$ && $SO(16)\times (SO(8))^2$ &\cr
height3pt&\omit      &&\omit             &&\omit            &\cr
& $E_8\times (SO(8))^2$ && $SO(16)\times (SU(2))^8$ && $SO(20)\times (SU
(2))^6$ &\cr
height3pt&\omit      &&\omit             &&\omit            &\cr
& $SO(16)\times (SO(8))^2$ && $(SO(8))^2\times (SU(2))^8$ && $SO(12)\times
SO(8)\times (SU(2))^6$ &\cr
height3pt&\omit      &&\omit             &&\omit            &\cr
& $(SO(8))^4$ && $(SU(2))^{16}$ && $SO(8)\times (SU(2))^{12}$ &\cr
height6pt&\omit      &&\omit             &&\omit            &\cr
\noalign{\hrule}
}}}

\endpage

\noindent
\centerline{Table 3.}
\centerline{Asymmetric ${\bf Z}_3$-orbifold models with (1,0) twisted
states.}
\vskip 4mm

\centerline{
\vbox{\offinterlineskip
\halign{&\vrule#&\strut\quad\hfil#\hfil\quad&\vrule#&\quad\hfil#\hfil
\quad&\vrule#&\quad\hfil#\hfil\quad&\vrule# \cr
\noalign{\hrule}
height3pt&\omit      &&\omit             &&\omit            &\cr
&  $G$                && $G_0$            && $G'$             &\cr
height3pt&\omit      &&\omit             &&\omit            &\cr
\noalign{\hrule}
height6pt&\omit      &&\omit             &&\omit            &\cr
& $D=6$\qquad\qquad\qquad\quad &&                &&                &\cr
height3pt&\omit      &&\omit             &&\omit            &\cr
& $E_6$              && $(SU(3))^3$      && $E_6$            &\cr
height3pt&\omit      &&\omit             &&\omit            &\cr
& $ (SU(3))^3$       && $(U(1))^6$       && $(SU(3))^3$      &\cr
height6pt&\omit      &&\omit             &&\omit            &\cr
& $D=12$\qquad\qquad\qquad\quad &&                &&                &\cr
height3pt&\omit      &&\omit             &&\omit            &\cr
& $(E_6)^2$    && $(SU(3))^6$      && $(E_6)^2$      &\cr
height6pt&\omit      &&\omit             &&\omit            &\cr
& $E_6\times (SU(3))^3$  && $(SU(3))^3\times (U(1))^6$  && $SU(6)\times SO
(8)\times (U(1))^3$      &\cr
height3pt&\omit      &&\omit             &&\omit            &\cr
& $ (SU(3))^6$       && $(U(1))^{12}$      && $(SU(2))^6\times (U(1))^6$
  &\cr
height3pt&\omit      &&\omit             &&\omit            &\cr
& $ E_8\times (SU(3))^2$   && $SU(9)\times (U(1))^4$   && $SO(20)\times (U
(1))^2$      &\cr
height6pt&\omit      &&\omit             &&\omit            &\cr
& $D=18$\qquad\qquad\qquad\  &&                &&                &\cr
height3pt&\omit      &&\omit             &&\omit            &\cr
& $(E_6)^3$    && $(SU(3))^9$    && $E_6\times (SU(3))^6$   &\cr
height3pt&\omit      &&\omit             &&\omit            &\cr
& $(E_6)^2\times (SU(3))^3$    && $(SU(3))^6\times (U(1))^6$
  && $SU(6)\times (SU(3))^4\times (U(1))^5$   &\cr
height3pt&\omit      &&\omit             &&\omit            &\cr
& $E_6\times (SU(3))^6$    && $(SU(3))^3\times (U(1))^{12}$
  && $SU(4)\times (SU(3))^2\times (U(1))^{11}$   &\cr
height3pt&\omit      &&\omit             &&\omit            &\cr
& $(SU(3))^9$        && $(U(1))^{18}$      && $SU(2)\times (U(1))^{17}$
&\cr
height3pt&\omit      &&\omit             &&\omit            &\cr
& $E_8\times E_6\times (SU(3))^2$    && $SU(9)\times (SU(3))^3\times (U
(1))^4$        && $SU(12)\times (SU(3))^2\times (U(1))^3$   &\cr
height3pt&\omit      &&\omit             &&\omit            &\cr
& $E_8\times (SU(3))^5$  && $SU(9)\times (U(1))^{10}$
  && $SU(10)\times (U(1))^9$   &\cr
height3pt&\omit      &&\omit             &&\omit            &\cr
& $(E_8)^2\times SU(3)$  && $(SU(9))^2\times (U(1))^2$
  && $SU(18)\times U(1)$   &\cr
height6pt&\omit      &&\omit             &&\omit            &\cr
\noalign{\hrule}
}}}

\endpage

\noindent
\centerline{Table 4.}
\centerline{Asymmetric ${\bf Z}_4$-orbifold models with (1,0) twisted
states.}
\vskip 4mm
\centerline{
\vbox{\offinterlineskip
\halign{&\vrule#&\strut\ \hfil#\hfil\ &\vrule#&\ \hfil#\hfil\ &\vrule#&\
\hfil#\hfil\ &\vrule# \cr
\noalign{\hrule}
height3pt&\omit      &&\omit             &&\omit            &\cr
&  $G$                && $G_0$            && $G'$             &\cr
height3pt&\omit      &&\omit             &&\omit            &\cr
\noalign{\hrule}
height6pt&\omit      &&\omit             &&\omit            &\cr
& $D=16$\qquad\qquad\quad\  &&                &&                &\cr
height3pt&\omit      &&\omit             &&\omit            &\cr
& $(E_8)^2$          && $(SO(10))^2\times (SU(4))^2$   && $(E_8)^2$   &\cr
height3pt&\omit      &&\omit             &&\omit            &\cr
& $SO(32)$           && $SU(8)\times (SO(8))^2\times U(1)$
  && $SO(24)\times SO(8)$ &\cr
height3pt&\omit      &&\omit             &&\omit            &\cr
& $SO(24)\times SO(8)$ && $SU(6)\times (SU(4))^2\times SU(2)\times (U
(1))^4$
  && $SU(12)\times SU(4)\times (U(1))^2$ &\cr
height3pt&\omit      &&\omit             &&\omit            &\cr
& $E_8\times SO(16)$ && $SO(10)\times (SU(4))^2\times (SU(2))^4\times U
(1)$
  && $(E_7)^2\times (SU(2))^2$ &\cr
height3pt&\omit      &&\omit             &&\omit            &\cr
& $(SO(16))^2$ && $(SU(4))^2\times (SU(2))^8 \times (U(1))^2$ && $(SO
(12))^2\times (SU(2))^4$  &\cr
height3pt&\omit      &&\omit             &&\omit            &\cr
& $E_8\times (SO(8))^2$ && $SO(10)\times SU(4)\times (SU(2))^2\times (U
(1))^6$    && $(E_6)^2\times (U(1))^4$ &\cr
height3pt&\omit      &&\omit             &&\omit            &\cr
& $SO(16)\times (SO(8))^2$ && $SU(4)\times (SU(2))^6\times (U(1))^7$
       && $(SU(6))^2\times (SU(2))^2\times (U(1))^4$ &\cr
height3pt&\omit      &&\omit             &&\omit            &\cr
& $(SO(8))^4$ && $(SU(2))^{4}\times (U(1))^{12}$  && $(SU(3))^4\times (U
(1))^8$ &\cr
height6pt&\omit      &&\omit             &&\omit            &\cr
\noalign{\hrule}
}}}

\endpage

\noindent
\centerline{Table 5.}
\centerline{Asymmetric ${\bf Z}_5$-orbifold models with (1,0) twisted
states.}
\vskip 4mm

\centerline{
\vbox{\offinterlineskip
\halign{&\vrule#&\strut\quad\hfil#\hfil\quad&\vrule#&\quad\hfil#\hfil
\quad&\vrule#&\quad\hfil#\hfil\quad&\vrule# \cr
\noalign{\hrule}
height3pt&\omit      &&\omit             &&\omit            &\cr
&  $G$                && $G_0$            && $G'$               &\cr
height3pt&\omit      &&\omit             &&\omit            &\cr
\noalign{\hrule}
height6pt&\omit      &&\omit             &&\omit            &\cr
& $D=4$\qquad\qquad\qquad\  &&                &&                &\cr
height3pt&\omit      &&\omit             &&\omit            &\cr
& $SU(5)$ && $(U(1))^4$                   && $SU(5)$            &\cr
height6pt&\omit      &&\omit             &&\omit            &\cr
& $D=8$\qquad\qquad\qquad\  &&                &&                &\cr
height3pt&\omit      &&\omit             &&\omit            &\cr
& $E_8$             && $(SU(5))^2$        && $E_8$              &\cr
height3pt&\omit      &&\omit             &&\omit            &\cr
& $(SU(5))^2$       && $(U(1))^8$         && $(SU(5))^2$        &\cr
height6pt&\omit      &&\omit             &&\omit            &\cr
& $D=12$\qquad\qquad\qquad\  &&                &&                &\cr
height3pt&\omit      &&\omit             &&\omit            &\cr
& $E_8\times SU(5)$   && $(SU(5))^2\times (U(1))^4$  && $SO(22)\times U
(1)$    &\cr
height3pt&\omit      &&\omit             &&\omit            &\cr
& $(SU(5))^3$       && $(U(1))^{12}$      && $(SU(4))^3\times (U(1))^3$
&\cr
height6pt&\omit      &&\omit             &&\omit            &\cr
& $D=16$\qquad\qquad\qquad\  &&                &&                &\cr
height3pt&\omit      &&\omit             &&\omit            &\cr
& $(E_8)^2$      && $(SU(5))^4$     && $(E_8)^2$    &\cr
height3pt&\omit      &&\omit             &&\omit            &\cr
& $E_8\times (SU(5))^2$  && $(SU(5))^2\times (U(1))^8$
      && $(SO(12))^2\times (U(1))^4$    &\cr
height3pt&\omit      &&\omit             &&\omit            &\cr
& $(SU(5))^4$ && $(U(1))^{16}$         && $(SU(2))^8\times (U(1))^8$   &
\cr
height6pt&\omit      &&\omit             &&\omit            &\cr
& $D=20$\qquad\qquad\qquad\  &&                &&                &\cr
height3pt&\omit      &&\omit             &&\omit            &\cr
& $(E_8)^2\times SU(5)$      && $(SU(5))^4\times (U(1))^4$
 && $(SU(10))^2\times (U(1))^2$    &\cr
height3pt&\omit      &&\omit             &&\omit            &\cr
& $E_8\times (SU(5))^3$  && $(SU(5))^2\times (U(1))^{12}$   && $(SU(6))^2
\times (U(1))^{10}$    &\cr
height3pt&\omit      &&\omit             &&\omit            &\cr
& $(SU(5))^5$ && $(U(1))^{20}$         && $(SU(2))^2\times (U(1))^{18}$
&\cr
height3pt&\omit      &&\omit             &&\omit            &\cr
\noalign{\hrule}
}}}

\endpage

\noindent
\centerline{Table 6.}
\centerline{Asymmetric ${\bf Z}_N$-orbifold models with (1,0) twisted
states.}
\vskip 4mm
\centerline{
\vbox{
\offinterlineskip
\halign{\vrule#&\strut\ \hfil#\ \hfil&\vrule#&\hfil\ #\hfil\ &\vrule#&
\hfil\ #\hfil\ &\vrule#&\hfil\ #\hfil\ &\vrule#\cr
\noalign{\hrule}
height2pt&\omit&&\omit&&\omit&&\omit&\cr
& ${\bf Z}_N$  && $G$          && $G_0$        && $G'$             &\cr
height2pt&\omit&&\omit&&\omit&&\omit&\cr
\noalign{\hrule}
height2pt&\omit&&\omit&&\omit&&\omit&\cr
& ${\bf Z}_6$  && $D=24$\qquad\qquad\quad   &&                &&
       &\cr
height3pt&\omit&&\omit&&\omit&&\omit&\cr
&        && $(E_8)^3$
         && $(SU(5))^3\times (SU(4))^3\times (U(1))^3$
         && $(SU(5))^6$                                      &\cr
height2pt&\omit&&\omit&&\omit&&\omit&\cr
& ${\bf Z}_7$  && $D=6$\qquad\qquad\quad   &&                &&
      &\cr
height3pt&\omit&&\omit&&\omit&&\omit&\cr
&        && $SU(7)$             && $(U(1))^6$     && $SU(7)$          &\cr
height2pt&\omit&&\omit&&\omit&&\omit&\cr
&        && $D=12$\qquad\qquad\quad   &&                &&
 &\cr
height3pt&\omit&&\omit&&\omit&&\omit&\cr
&        && $(SU(7))^2$         && $(U(1))^{12}$   && $(SU(6))^2\times (U
(1))^2$  &\cr
height2pt&\omit&&\omit&&\omit&&\omit&\cr
&        && $D=18$\qquad\qquad\quad   &&                &&
 &\cr
height3pt&\omit&&\omit&&\omit&&\omit&\cr
&        && $(SU(7))^3$         && $(U(1))^{18}$   && $(SU(2))^9\times (U
(1))^9$  &\cr
height2pt&\omit&&\omit&&\omit&&\omit&\cr
& ${\bf Z}_9$  && $D=18$\qquad\qquad\quad   &&                &&
       &\cr
height3pt&\omit&&\omit&&\omit&&\omit&\cr
&        && $(E_6)^3$
         && $(SU(2))^3\times (U(1))^{15}$
         && $(SO(8))^3\times (U(1))^6$                                 &
\cr
height2pt&\omit&&\omit&&\omit&&\omit&\cr
& ${\bf Z}_{10}$  && $D=8$\qquad\qquad\quad   &&                &&
         &\cr
height3pt&\omit&&\omit&&\omit&&\omit&\cr
&        && $E_8$     && $(SU(3))^2\times (SU(2))^2\times (U(1))^2$   &&
$E_8$     &\cr
height2pt&\omit&&\omit&&\omit&&\omit&\cr
&        && $D=16$\qquad\qquad\quad   &&                &&
 &\cr
height3pt&\omit&&\omit&&\omit&&\omit&\cr
&        && $(E_8)^2$
         && $(SU(3))^4\times (SU(2))^4\times (U(1))^4$
         && $SO(32)$                                                &\cr
height2pt&\omit&&\omit&&\omit&&\omit&\cr
&        && $D=24$\qquad\qquad\quad   &&             &&                  &
\cr
height3pt&\omit&&\omit&&\omit&&\omit&\cr
&        && $(E_8)^3$
         && $(SU(3))^6\times (SU(2))^6\times (U(1))^6$
         && $(SU(3))^{12}$
&\cr
height2pt&\omit&&\omit&&\omit&&\omit&\cr
& ${\bf Z}_{11}$  && $D=10$\qquad\qquad\quad   &&          &&
    &\cr
height3pt&\omit&&\omit&&\omit&&\omit&\cr
&        && $SU(11)$     && $(U(1))^{10}$       && $SU(11)$         &\cr
height2pt&\omit&&\omit&&\omit&&\omit&\cr
&        && $D=20$\qquad\qquad\quad   &&             &&                  &
\cr
&        && $(SU(11))^2$
         && $(U(1))^{20}$
         && $(SU(2))^{10}\times(U(1))^{10}$                         &\cr
height2pt&\omit&&\omit&&\omit&&\omit&\cr
& ${\bf Z}_{13}$  && $D=12$\qquad\qquad\quad   &&          &&
    &\cr
height3pt&\omit&&\omit&&\omit&&\omit&\cr
&        && $SU(13)$     && $(U(1))^{12}$       && $SU(12)\times U(1)$ &
\cr
height2pt&\omit&&\omit&&\omit&&\omit&\cr
& ${\bf Z}_{15}$  && $D=24$\qquad\qquad\quad   &&           &&
     &\cr
height3pt&\omit&&\omit&&\omit&&\omit&\cr
&        && $(E_8)^3$
         && $(SU(2))^{12}\times (U(1))^{12}$
         && $(SU(2))^{24}$                                          &\cr
height2pt&\omit&&\omit&&\omit&&\omit&\cr
& ${\bf Z}_{17}$  && $D=16$\qquad\qquad\quad   &&          &&
    &\cr
height3pt&\omit&&\omit&&\omit&&\omit&\cr
&        && $SU(17)$     && $(U(1))^{16}$    && $(SU(8))^2\times (U(1))^2$
&\cr
height2pt&\omit&&\omit&&\omit&&\omit&\cr
& ${\bf Z}_{19}$  && $D=18$\qquad\qquad\quad   &&          &&
    &\cr
height3pt&\omit&&\omit&&\omit&&\omit&\cr
&        && $SU(19)$     && $(U(1))^{18}$    && $(SU(6))^3\times (U(1))^3$
&\cr
height2pt&\omit&&\omit&&\omit&&\omit&\cr
& ${\bf Z}_{20}$  && $D=16$\qquad\qquad\quad   &&           &&
     &\cr
height3pt&\omit&&\omit&&\omit&&\omit&\cr
&        && $(E_8)^2$
         && $(SU(2))^4\times (U(1))^{12}$
         && $(E_8)^2$                                          &\cr
height2pt&\omit&&\omit&&\omit&&\omit&\cr
& ${\bf Z}_{23}$  && $D=22$\qquad\qquad\quad   &&          &&
    &\cr
height3pt&\omit&&\omit&&\omit&&\omit&\cr
&  && $SU(23)$     && $(U(1))^{22}$    && $(SU(2))^{11}\times (U(1))^{11}$
&\cr
height2pt&\omit&&\omit&&\omit&&\omit&\cr
& ${\bf Z}_{30}$  && $D=24$\qquad\qquad\quad   &&           &&
     &\cr
height3pt&\omit&&\omit&&\omit&&\omit&\cr
&        && $(E_8)^3$  && $(U(1))^{24}$   && $(U(1))^{24}$ &\cr
height2pt&\omit&&\omit&&\omit&&\omit&\cr
\noalign{\hrule}
}}}

\end